\documentclass{aa}
\usepackage{rotating}
\usepackage{graphicx}
\usepackage{times}

\font\twlbfs=cmbxsl10 scaled \magstep3

\begin{document}

\author{I.I.~Antokhin\inst{1,2,3}
        \and G.~Rauw\inst{2}\thanks{Research Associate FNRS (Belgium).}
        \and J.-M.~Vreux\inst{2}
        \and K.A.~van der Hucht\inst{4,5}
        \and J.C.~Brown\inst{3}
       }
\institute{Sternberg Astronomical Institute, Moscow
           University, Universitetskij Prospect 13, Moscow 119992, Russia
 \and      Institut d'Astrophysique et de G\'eophysique, Universit\'e de
           Li\`ege, All\'ee du 6 Ao\^ut, 17 B\^at. B5c, B-4000
           Li\`ege, Belgium
 \and      Department of Physics and Astronomy, University of Glasgow,
           Kelvin Building, Glasgow G12\,8QQ, Scotland, UK
 \and      SRON Netherlands Institute for Space Research, Sorbonnelaan 2,
           NL-3584\,CA Utrecht, the Netherlands
 \and      Astronomical Institute Anton Pannekoek, University of Amsterdam,
           Kruislaan 403, NL-1098\,SJ Amsterdam, the Netherlands
          }

\title {
         {\twlbfs XMM-Newton} X-ray study of early type stars in \\
         the Carina\, OB1 association \thanks{Based on observations
         obtained with {\sl XMM-Newton}, an ESA science mission with
         instruments and contributions directly funded by ESA Member
         States and the USA (NASA). The X-ray catalogue and its
         cross-identification with infra-red and optical catalogues
         (Tables 2 and 3) are only available in electronic form at
         the CDS via anonymous ftp to cdsarc.u-strasbg.fr
         (130.79.128.5) or via
         http://cdsweb.u-strasbg.fr/cgi-bin/qcat?J/A+A/.
       }
       }

\authorrunning{I.I.~Antokhin et al.}
\titlerunning{{\sl XMM} study of early type stars}

\offprints {I.I.~Antokhin \email{ igor@sai.msu.ru} }

\date{Received date \today; accepted date $\ldots$}

\abstract
{}
{X-ray properties of the stellar population in the Carina\, OB1
association are examined with special emphasis on early-type stars. Their
spectral characteristics provide some clues to understanding the nature of
X-ray formation mechanisms in the winds of single and binary early-type
stars.}
{A timing and spectral analysis of five observations with XMM-Newton is
performed using various statistical tests and thermal spectral models.}
{235 point sources have been detected within the field of
view. Several of these sources are probably pre-main sequence stars
with characteristic short-term variability. Seven sources are
possible background AGNs. Spectral analysis of twenty three sources
of type OB and WR\,25 was performed. We derived spectral parameters
of the sources and their fluxes in three energy bands. Estimating the
interstellar absorption for every source and the distance to the
nebula, we derived X-ray luminosities of these stars and compared
them to their bolometric luminosities. We discuss possible reasons
for the fact that, on average, the observed X-ray properties of
binary and single early type stars are not very different, and give
several possible explanations.}
{}


\maketitle


\section{Introduction}

Although \mbox{X-ray} emission by early-type stars is nowadays well
established, its origin is still not fully understood. Observations
with the {\sl Einstein}, {\sl ROSAT} and {\sl ASCA} satellites
indicated that hot luminous stars have rather soft thermal {\mbox
X-ray} spectra. This picture contrasts with the expected spectral
properties if the {\sl X-rays} were produced in a hot corona at the
base of the stellar wind (e.g.\ Waldron \cite{Waldr84}). Indeed, in
the latter case one would expect to observe substantial absorption
of the softest emission. The lack of this absorption in the observed
spectra triggered the elaboration of an alternative scenario. In
fact, according to the phenomenological model proposed by Lucy \&
White (\cite{Lucy80}) and further elaborated by Lucy
(\cite{Lucy82}), hydrodynamic shocks are generated throughout a
radiation driven stellar wind as the consequence of the intrinsic
instabilities of radiative driving. The velocity jumps in such
shocks heat the post-shock plasma to temperatures of a few million
degrees. These shocks would be distributed throughout the stellar
wind with the result that even soft {\mbox X-rays} could escape the
wind without substantial absorption. Recent theoretical studies of
such instabilities have been presented for instance by Feldmeier et
al.\ (\cite{Feld97}), Owocki \& Cohen (\cite{OwockiC99}), and
Dessart \& Owocki (\cite{Dess02}). High resolution spectra obtained
with the new generation of {\mbox X-ray} observatories have
confirmed the thermal origin of the bulk of the {\mbox X-ray}
emission from early-type stars (e.g.\ Kahn et al.\ \cite{Kahn01}),
but in some cases, the properties of the observed line profiles led
also to new challenges for the shock model (see e.g.\ Rauw
\cite{ElEscorial} for a review).

Another well established, though yet unexplained, property of the {\mbox
X-ray} emission of hot stars is the linear scaling between the
observed {\mbox X-ray} luminosity ($L_{\rm X}$) and the bolometric luminosity
($L_ {\rm bol}$). Using {\sl Einstein} data, Long \& White
(\cite{Long80}), Pallavicini et al.\ (\cite{Pall81}) and Chlebowski et al.\
(\cite{Chle89}) found that $L_{\rm X} \simeq 10^{-7} \times L_{\rm bol}$ for OB stars.
{\sl ROSAT} observations broadly confirmed this relation. Bergh\"ofer et
al.\ (\cite{Berg97}) and Kudritzki et al.\ (\cite{Kudrit96}) found
that including a weak dependence on the characteristic wind density leads
to a somewhat tighter relationship. However, considering the scaling
of the X-ray emission with various wind and stellar properties
in the framework of the shock model, Owocki \& Cohen (\cite{OwockiC99})
found that a delicate equilibrium between emission and absorption is required to
reproduce the empirically derived $L_{\rm X}$ -- $L_{\rm bol}$ relation.
With the broader energy range and improved sensitivity of the new
generation of {\mbox X-ray} satellites it is important to re-address the
question of the empirical $L_{\rm X}$ -- $L_{\rm bol}$ scaling for OB stars.

When doing this it is crucial to pay attention to the multiplicity
of the stars in the sample. Indeed, in OB binaries, the collision
of the two stellar winds is expected to generate a strong {\mbox
X-ray} bright shock between the stars (see e.g.\ Stevens et al.\
\cite{Stev92}) and this feature is believed to manifest itself as an
``excess'' in the $L_{\rm X}/L_{\rm bol}$ ratio (see e.g. Pollock\
\cite{Pollock95}). Moreover, since the relative velocities of colliding
winds can be much higher than the shock velocity jumps in winds of single
stars, the {\mbox X-ray} spectra of colliding wind early type binaries can
be quite hard. In fact the hardness of the observed spectra of some
early-type stars, and, in particular, the presence of the Fe\,XXV and
Fe\,XXVI lines at $~6.7$\,keV indicative of very hot material has been
suggested as an indication of possible binarity (e.g. Raassen et al.,
\cite{Raas03}).

This latter suggestion leads to an interesting question. Clearly in a
general case it is very difficult to ``hide'' such a hard secondary
component of a colliding wind binary. Indeed, to produce strong hard
\mbox{X-ray} emission, the companion must be a massive hot star, and as
such, should manifest itself in other wavelength domains (e.g.\ optical
spectra) and also via orbital spectral and photometric variability in the
\mbox{X-rays}. Traditionally, a lack of variations was attributed to a
``pole-on'' orientation of the binary orbit. However, lately, a growing
number of examples of apparently single O and WR stars with hard {\mbox
X-ray} spectra (e.g. Skinner et al.\ \cite{Skinn02}, Raassen et al.
\cite{Raas03}, the present study) has become available. It seems
unlikely that all these objects are binaries seen from their orbital poles.
Thus the question is whether there exists another, yet unknown, mechanism
which could produce hard {\mbox X-rays} in single star winds? Exceptions to
this rule are stars such as $\theta^1$\,Ori\,C with strong enough magnetic
fields to confine the stellar wind into the plane near the magnetic equator
(Cassinelli et al., \cite{Cassin02}, Brown et al., \cite{Brown03}, Maheswaran
\cite{Mahes03}, Gagn\'e et al.\ \cite{Gagne}, Li et al., \cite{Li07}). In
such cases, the head-on collision of the winds from the two hemispheres
of a single star produces a rather hard X-ray emission. The fundamental
problem here is that the only source of energy in the classical picture of a
hot star wind which is able to produce high temperatures and hence hard
thermal {\mbox X-rays}, is the wind kinetic energy. But to release most of
this energy through radiation one needs a ``wall'' able to stop the
supersonic flow and heat the material in the formed shock. In the winds of
single stars we do not expect to find such a wall except in the magnetised
case. On the other hand, there is a number of known early-type binaries
which show neither significant X-ray excess nor particularly hard X-ray
spectra. What could be the mechanism which dumps the X-ray production in
these systems?

The Carina OB\,1 association with its impressive number of
early-type stars -- including some of the youngest, hottest and most
massive objects known in our Galaxy -- is an ideal place to
investigate the {\mbox X-ray} properties of these objects. The
history of X-ray studies of the region goes back to the {\em
Einstein} era. Seward et al. (\cite{Seward79}) presented {\em
Einstein} X-ray (0.2\,--\,4.0\,keV) observations of the Carina open
cluster Tr\,16 and its environment, including six O-type stars and
one WR star, WR\,25. Subsequent {\em Einstein} observations by
Seward \& Chlebowski (\cite{Seward82}) of the same region showed
X-rays from 15 O-type and WR stars. More studies were performed with
the use of the {\em ROSAT} and {\em ASCA} satellites. Some of them
were focused on individual objects like {\em the LBV} $\eta$\,Car
(Corcoran et al. \cite{Corcoran00} and references therein) or WR\,25
(e.g. Pollock \& Corcoran \cite{Pollock06}, Skinner et al.
\cite{Skinner95}). Others were devoted to investigating general
X-ray properties of the early-type stars population (Corcoran
\cite{Corc99} and references therein).

Recently, four studies of the region based on X-ray observations
with {\em Chandra} and {\sl XMM-Newton} satellites were published.
Evans et al. (\cite{Evans03}) (hereafter EV03) and Evans et al.
(\cite{Evans04}) (hereafter {\em EV04}) using {\em Chandra} data,
detected 154 point sources, 23 of which are of O, B types. They
present luminosities and hardness ratios for the detected sources
and confirm the $L_{\rm X}$ -- $L_{\rm bol}$ relation for early-type
stars. They also discuss the low-resolution spectra of the 14
brightest sources. Sanchawala et al. (\cite{Sanchawala07}), using
the same {\em Chandra} data as EV03 plus another archival {\em
Chandra} data set, detected 454 sources in the region, among which
38 are known OB stars. The latter paper is focused on
cross-identification of the X-ray sources with optical and
near-infrared imaging observations specifically obtained for this
study. This allowed the authors to conclude that about 300 sources
are likely late-type pre-main-sequence stars. The X-ray luminosities
of the detected OB stars were estimated based on their count rate
and using a single-temperature Raymond-Smith model of thermal plasma
with $\log\,T=6.65$ (0.384\,keV). Albacete-Colombo et al.
(\cite{AC03}) (hereafter AC03) detected 80 discrete sources in the
region using two observations by {\sl XMM-Newton} (revolution
numbers 115 and 116). They discuss X-ray properties of the sources,
including the $L_{\rm X}$ -- $L_{\rm bol}$ relation for early-type
stars. In the present paper, we used three additional {\sl
XMM-Newton} observations, which allow us to increase the signal to
noise ratio and to detect more than 200 discrete sources. We shall
compare our results with those of AC03 throughout the current paper.
We do not present a detailed spectral analysis of individual
objects in this paper as many such objects deserve dedicated
papers. A detailed study of WR\,25 based on the same {\sl
XMM-Newton} data as in the current study was published by Raassen et
al. (\cite{Raas03}), whilst $\eta$\,Car has been the subject of many
studies including those utilizing extensive data sets obtained with
various X-ray observatories (see, e.g., Leutenegger et al.
\cite{Leut03}; Weis et al. \cite{Weis04}; Viotti et al.
\cite{Viotti04}; Corcoran \cite{Corc05}; Hamaguchi et al.
\cite{Hamaguchi07}). Analysis of other interesting objects (e.g.
the early type binary HD\,93205) will be the subject of forthcoming
papers.

The goals of this paper are: \vspace*{-2mm}
\begin{itemize}
\item To describe general X-ray properties of the detected sources.
\item To look for their variability, both short and long term.
\item To investigate the $L_{\rm X}/L_{\rm bol}$ relation for early type stars.
\item To compare X-ray properties of single versus binary early-type stars
and to try to get some clues to their differences and similarities.
\end{itemize}

The structure of the paper follows these outlined priorities. In the
second section, the observational details are given. Source
detection, their identification with the optical and IR catalogues,
X-ray count rates, identification of possible extra-galactic sources
are given in Section 3. Section 4 provides general
information about open clusters in the Carina nebula, along with the
discussion of their distance from the Earth. The results of the
variability study are presented in Section 5. The X-ray
properties of early-type stars detected in the region are presented
in Section 6. We discuss our results in section 7
and a short summary is given in Section 8.

\section{Observations}

The observations of the Carina region were obtained with the {\sc
r}eflection {\sc g}rating {\sc s}pectrometers ({\sc rgs}) and the {\sc
e}uropean {\sc p}hoton {\sc i}maging {\sc c}amera ({\sc epic}) {\sc ccd}
detectors of the {\sl XMM-Newton} observatory. The log of {\sc epic}
observations is shown in Table~\ref{obslog}. In the two first data sets the
primary target was $\eta$\,Car while in the third to fifth data sets the
primary target was WR\,25. The analysis of {\sc rgs} and {\sc epic}
observations of WR\,25 was presented in Raassen et al. (\cite{Raas03}) while
the {\sc RGS} analysis of $\eta$\,Car was published by Leutenegger et al.
(\cite{Leut03}) and we will not repeat these here. In
Fig.~\ref{plot_coords} we show the mozaiced image of the area combined from
all 5 data sets and 3 {\sc epic} instruments.

\begin{table}
\caption{Log of the Carina region observations by {\sl XMM-Newton}.}
\label{obslog}
    \leavevmode
\setlength{\tabcolsep}{0.4mm}
\begin{tabular}[h]{l|ccccc}
\hline\hline
                 &          &          &          &          &          \\[-2mm]
Data Set Number  &  1       &    2     &    3     &    4     &     5    \\
Revolution       & \multicolumn{1}{c}{ \#\,115}
                 & \multicolumn{1}{c}{ \#\,116}
                 & \multicolumn{1}{c}{ \#\,283}
                 & \multicolumn{1}{c}{ \#\,284}
                 & \multicolumn{1}{c}{ \#\,285}                         \\
Obs. Date        & \multicolumn{1}{c}{26-7-2000}
                 & \multicolumn{1}{c}{27-7-2000}
                 & \multicolumn{1}{c}{25-6-2001}
                 & \multicolumn{1}{c}{28-6-2001}
                 & \multicolumn{1}{c}{30-6-2001}                        \\
                 &          &          &          &          &          \\[-2mm]
\hline
                 &          &          &          &          &          \\[-2mm]
Start (UT)       & 04:58    & 23:48    & 06:51    & 07:22    & 04:38    \\
                 &          &          &          &          &          \\[-2mm]
\hline\hline
                 &          &          &          &          &          \\[-2mm]
Instrument       & \multicolumn{5}{c}{Integration Time}            \\
                 & \multicolumn{5}{c}{(hr)}            \\
                 &          &          &          &          &          \\[-2mm]
\hline
                 &          &          &          &          &          \\[-2mm]
{\sc mos\,1}     & 9.4      & 3.1      & 10.2     & 11.7     & 10.4     \\
{\sc mos\,2}     & 8.5      & 2.3      & 10.2     & 11.7     & 10.4     \\
{\sc pn}         & 8.8      & 2.6      & ~9.6     & 11.0     & ~9.7     \\[-2mm]
                 &          &          &          &          &          \\
\hline\hline
\end{tabular}
\end{table}

The {\sc epic-mos} and {\sc pn} instruments were operated in the full frame
mode except during rev. 115 and 116 where {\sc mos2} was operated in the
small window mode. All three {\sc epic} instruments used the thick filter to
reject optical light. We used version 7.0 of the {\sl XMM-Newton} Science
Analysis System (SAS) software to reduce the raw {\sc epic} data. For
spectral analyses, we adopted the most up-to-date redistribution matrices
provided by the {\sc epic} instrument teams and used SAS to build the
appropriate ancillary response file for each observation. More details on
the pipeline processing of the data are given in Raassen et al.
(\cite{Raas03}).


\section{Source detection and cross-identification}

\subsection{Source detection}

Sources were detected and count rates measured in three energy
bands -- soft (0.4\,-\,1.0\,keV), medium (1.0\,-\,2.5\,keV), and hard
(2.5\,-\,10.0\,keV). The choice of these bands is based on the following
considerations:

\begin{itemize}

\item There is a lot of noise (sometimes systematic features) in the {\sc pn}
data for $E\,<\,0.4$\,keV.

\item The {\sc epic} sensitivity above $10$\,keV is almost zero.

\item The relatively narrow soft band is sensitive to interstellar
absorption.

\item The wide hard band allows one to collect more photons and hence
improve statistics.

\end{itemize}

Source detection and determination of source parameters was
performed with the SAS {\tt edetect\_chain} metatask based on the
{\em sliding cell detection} and {\em maximum likelihood} (ML)
methods. The ML method (the SAS task {\tt emldetect}) makes use
of maximum likelihood PSF (point spread function) fitting to the
source count distribution. The so-called logarithmic detection
likelihoods obey the simple relationship $L_2 = -\ln(p)$ where $p$
is the probability for a random Poissonian fluctuation to have
caused the observed source counts. The threshold for inclusion of
sources in the final output was set to $L_2^{min}=10.0$. In the
limit of a very large number of counts, the likelihood function is a
gaussian and the logarithmic likelihood becomes a parabola. At this
limit the value of $10.0$ is equivalent to 3 sigmas.

To increase the signal-to-noise ratio and thus to detect as
many sources as possible, we merged the event lists and images from
different data sets. Unfortunately, the angular distance between the
central axes of the data sets 1 and 2 on one hand, and 3 to 5 on the
other, is too large ($\sim\,7\arcmin$) and does not allow one to
merge all data sets into a single coordinate system without
producing spurious results in the subsequent source detection. For
this reason we merged the data sets 1 and 2, and the data sets 3, 4,
5 separately. The source detection was performed on the two
resulting merged sets of data, making use of all energy bands and
all instruments simultaneously. The resulting source lists were
visually inspected and a few spurious detections, mainly along the
CCD edges, were removed.

The coordinates of the sources common in the two merged data sets
are very close and do not show any systematic differences. We
attempted to use the {\tt eposcorr} procedure to improve the
coordinates by cross-correlating the XMM coordinates with the Two
Micron All Sky Survey point source catalogue\footnote{See
http://irsa.ipac.caltech.edu/} (2MASS, Skrutskie et al.\
\cite{2mass}) and the optical sources in the SIMBAD database. In
both cases, neither a significant systematic shift nor a field
rotation was apparent. The corrections in RA and DEC were less than
0.5\arcsec and not consistent between the IR and optical catalogues.
We conclude that the procedure does not allow one to improve the
{\sl XMM-Newton} coordinates. For the final source list the
coordinates of the sources common in the two merged data sets were
averaged. As some of the detected sources may be variable, it is of
interest to measure their count rates not just in the merged data
sets but in every individual one. For this purpose, we repeated the
ML routine with our source list as the input, for the data sets 1 --
5.

\begin{figure*}
\centering
\includegraphics[width=15.0truecm]{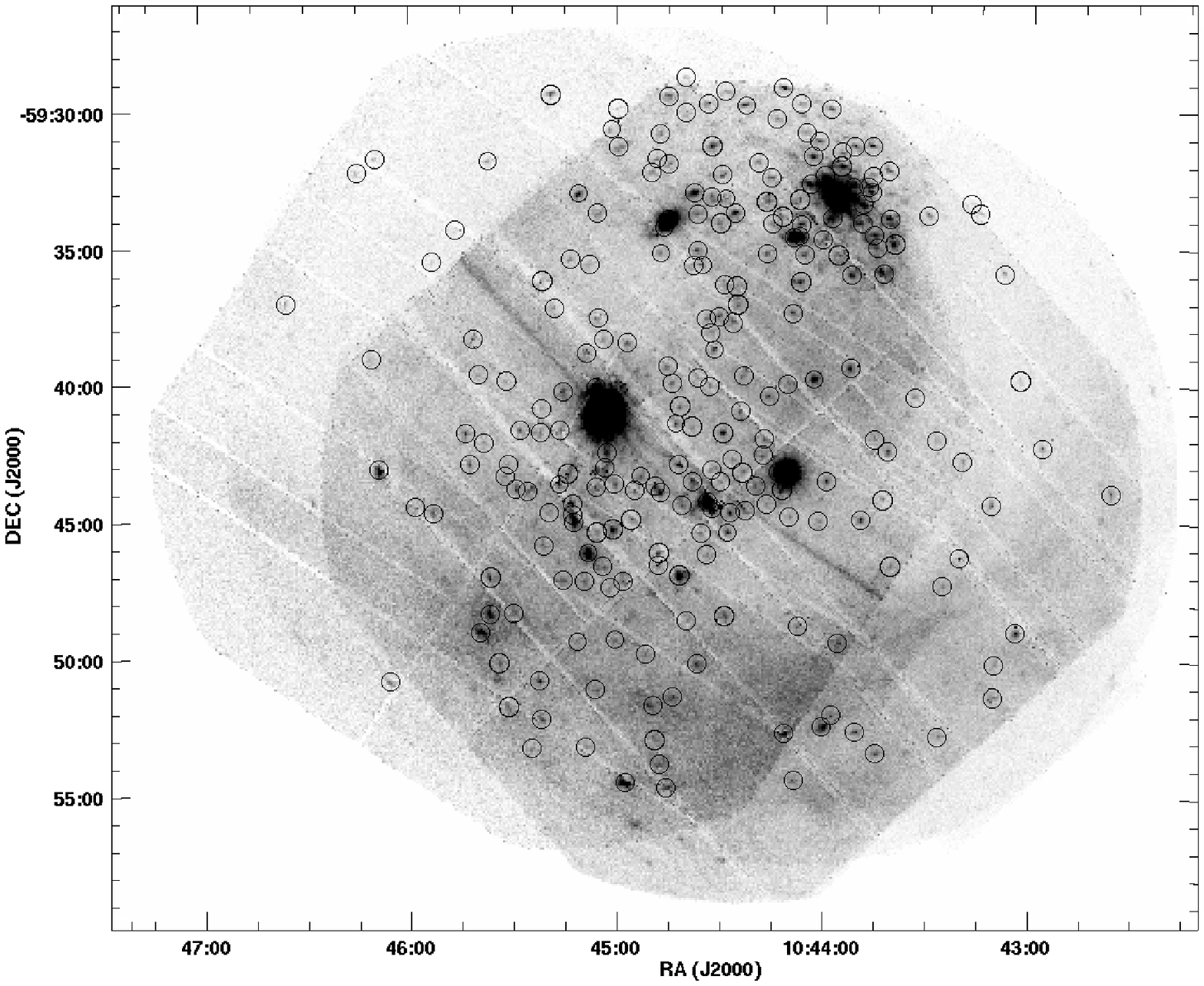}

\caption{Mosaiced exposure-corrected image of the field, combining
all data sets and all instruments {\sc mos1}, {\sc mos2}, and {\sc pn}.
Energy range 0.4 -- 10.0keV. Positions of the detected sources are shown.}

\label{plot_coords}
\end{figure*}

The total number of detected sources is 235. Recall that AC03,
using the data sets 1 and 2, detected 80 sources. Sanchawala et al.
(\cite{Sanchawala07}) were able to detect 454 sources in the {\em
Chandra} data on the field, thanks to the superior angular
resolution of the {\em Chandra} telescope. Fig.\ref{plot_coords}
shows the positions of the detected sources over the mozaiced image
of the field. A sample of the catalogue of the detected
discrete X-ray sources is presented in Table\,\ref{table_cts}. The
full table is available as online data. The table includes only
those measurements in which the total logarithmic {\sc epic}
likelihood exceeds the value of $10.0$.

One should note that the count rates in Table\,\ref{table_cts} may
be systematically different from those provided for the same stars
by AC03. One evident reason is that we used different energy limits
for our soft, medium, and hard bands than AC03. A deeper reason, is,
however, that AC03 used the wavelet technique (the SAS {\tt
ewavelet} routine) to estimate their count rates. Using {\tt
ewavelet} is equivalent to fit a Gaussian to the observed discrete
source image. However, the actual {\sc epic} PSF is quite different
from a Gaussian. The {\tt emldetect} routine makes use of an
empirical PSF and thus, according to the SAS manual, should provide
more accurate count rate estimates than {\tt ewavelet}. Our tests
show that the AC03 count rates are systematically lower than the
count rates obtained for the same energy band with {\tt emldetect}.

\begin{table*}

\caption{Sample of the X-ray catalogue.}

\label{table_cts}
\begin{tabular}{ rcr|rrrrrrrrrr }
\hline
 X\# & Var Status & Data Set & $L_2^{\rm PN}$ & $cr_{\rm PN}$ & $\sigma_{\rm PN}$ & $cr_{\rm PN}^{\rm S}$ & $\sigma_{\rm PN}^{\rm S}$ & $cr_{\rm PN}^{\rm M}$ & $\sigma_{\rm PN}^{\rm M}$ & $cr_{\rm PN}^{\rm H}$ & $\sigma_{\rm PN}^{\rm H}$ \\
$[1]$ &  $[2]$  &   $[3]$ &   $[4]$   &   $[5]$   &   $[6]$   &  $[7]$    &  $[8]$    &  $[9]$    &  $[10]$   & $[11]$    &  $[12]$  \\[+1mm]
\hline
  1 &   Uncert  &     3   &     37.0  &  11.6964  &   1.8388  &   7.5714  &   1.5446  &   4.1250  &   0.9757  &   0.0000  &   0.2079  \\
    &           &     5   &     67.8  &  14.3821  &   1.6988  &   9.8716  &   1.4067  &   3.6847  &   0.7888  &   0.8258  &   0.5338  \\
  2 &   Const   &     3   &     78.9  &  11.0794  &   1.5107  &   1.0840  &   0.7725  &   4.9085  &   0.8522  &   5.0868  &   0.9793  \\
    &           &     5   &    155.1  &  13.0312  &   1.2820  &   0.0000  &   0.3180  &   5.0587  &   0.7272  &   7.9724  &   1.0067  \\
  3 &   Noinfo  &     3   &      5.9  &   1.5988  &   0.6920  &   0.0000  &   0.3058  &   0.0631  &   0.2185  &   1.5357  &   0.5810  \\
  4 &   Var     &     3   &     14.2  &  13.1136  &   3.7333  &   3.0763  &   2.6452  &   6.8288  &   2.0792  &   3.2085  &   1.6178  \\
    &           &     4   &    257.1  &  30.2308  &   2.6712  &   6.3221  &   1.5744  &  19.5985  &   1.8292  &   4.3102  &   1.1448  \\
    &           &     5   &    165.1  &  18.3160  &   1.7907  &   5.1892  &   1.2056  &  10.6765  &   1.1067  &   2.4504  &   0.7269  \\
  5 &   Uncert  &     3   &     27.6  &   6.2785  &   1.2448  &   4.3118  &   0.9343  &   1.2414  &   0.5292  &   0.7254  &   0.6297  \\
    &           &     4   &     46.6  &   8.9067  &   1.4053  &   5.4440  &   1.0365  &   2.8946  &   0.7262  &   0.5681  &   0.6108  \\
    &           &     5   &     25.8  &   4.7296  &   0.9220  &   3.1683  &   0.7315  &   1.3936  &   0.4500  &   0.1676  &   0.3355  \\
  6 &   Const   &     3   &     41.6  &   8.5184  &   1.6391  &   0.7818  &   0.9702  &   2.2362  &   0.7546  &   5.5004  &   1.0844  \\
    &           &     5   &     59.4  &   8.3989  &   1.3907  &   0.0000  &   0.7206  &   2.0239  &   0.6175  &   6.3750  &   1.0165  \\
  7 &   Const   &     3   &     27.9  &   6.7066  &   1.7137  &   0.9344  &   1.1214  &   5.7722  &   1.1112  &   0.0000  &   0.6666  \\
    &           &     4   &      3.0  &   2.7511  &   1.4317  &   0.7435  &   0.9801  &   2.0076  &   0.8730  &   0.0000  &   0.5718  \\
  8 &   Const   &     3   &     64.3  &  11.2433  &   1.4168  &   7.0934  &   1.1558  &   4.1499  &   0.7856  &   0.0000  &   0.2330  \\
    &           &     4   &     81.9  &  11.4430  &   1.3621  &   6.0041  &   1.0333  &   5.4389  &   0.8331  &   0.0000  &   0.3058  \\
    &           &     5   &     89.4  &  11.0901  &   1.2253  &   7.4002  &   1.0139  &   3.6899  &   0.6185  &   0.0000  &   0.3015  \\
\hline
\hline
 X\# & Var Status & Data Set & $L_2^{\rm MOS1}$ & $cr_{\rm MOS1}$ & $\sigma_{\rm MOS1}$ & $cr_{\rm MOS1}^{\rm S}$ & $\sigma_{\rm MOS1}^{\rm S}$ & $cr_{\rm MOS1}^{\rm M}$ & $\sigma_{\rm MOS1}^{\rm M}$ & $cr_{\rm MOS1}^{\rm H}$ & $\sigma_{\rm MOS1}^{\rm H}$ \\
$[1]$ &  $[2]$  &  $[3]$  &  $[13]$  &   $[14]$  &  $[15]$   & $[16]$    & $[17]$    & $[18]$    &  $[19]$   & $[20]$    &  $[21]$  \\[+1mm]
\hline
  1 &   Uncert  &     3   &     2.8  &   1.5736  &   0.7445  &   0.2249  &   0.3832  &   1.1755  &   0.5769  &   0.1732  &   0.2732  \\
    &           &     5   &    30.0  &   4.5064  &   0.8380  &   2.3016  &   0.6342  &   2.2048  &   0.5365  &   0.0000  &   0.1103  \\
  2 &   Const   &     3   &    41.7  &   3.7951  &   0.7415  &   0.0000  &   0.1756  &   2.1814  &   0.5132  &   1.6137  &   0.5057  \\
    &           &     5   &    54.6  &   4.0763  &   0.6695  &   0.1027  &   0.2009  &   1.7207  &   0.3944  &   2.2529  &   0.5023  \\
  3 &   Noinfo  &     3   &     6.3  &   1.1639  &   0.4805  &   0.0000  &   0.1250  &   0.4476  &   0.3061  &   0.7163  &   0.3486  \\
  4 &   Var     &     3   &    26.8  &   5.4286  &   1.0912  &   1.3413  &   0.6648  &   3.0680  &   0.7086  &   1.0192  &   0.4967  \\
    &           &     4   &   116.8  &  11.9558  &   1.3586  &   2.6415  &   0.7379  &   6.3675  &   0.8938  &   2.9468  &   0.7088  \\
    &           &     5   &    66.2  &   7.3132  &   1.0267  &   1.9323  &   0.6414  &   3.8654  &   0.6433  &   1.5155  &   0.4785  \\
  5 &   Uncert  &     3   &    10.7  &   1.8870  &   0.6134  &   0.7426  &   0.3567  &   0.8337  &   0.3534  &   0.3107  &   0.3524  \\
    &           &     4   &    25.3  &   3.5911  &   0.8622  &   1.1051  &   0.4992  &   2.4859  &   0.6274  &   0.0000  &   0.3171  \\
    &           &     5   &     6.5  &   1.2127  &   0.4765  &   1.0220  &   0.4014  &   0.1907  &   0.2387  &   0.0000  &   0.0948  \\
  6 &   Const   &     3   &     8.0  &   1.8536  &   0.7433  &   0.0000  &   0.3138  &   0.2292  &   0.3111  &   1.6244  &   0.5977  \\
    &           &     5   &    15.4  &   2.5796  &   0.7105  &   0.1040  &   0.2827  &   0.8577  &   0.4188  &   1.6178  &   0.4994  \\
  7 &   Const   &     3   &     3.8  &   1.6451  &   0.7503  &   0.2117  &   0.4138  &   0.9603  &   0.4802  &   0.4731  &   0.4013  \\
    &           &     4   &     1.5  &   0.9795  &   0.6168  &   0.0000  &   0.2356  &   0.6828  &   0.4627  &   0.2968  &   0.3328  \\
  8 &   Const   &     3   &    26.0  &   3.5804  &   0.7153  &   1.7610  &   0.5240  &   1.7545  &   0.4686  &   0.0648  &   0.1322  \\
    &           &     4   &    39.5  &   4.2651  &   0.7098  &   2.9362  &   0.5725  &   1.0653  &   0.3545  &   0.2635  &   0.2246  \\
    &           &     5   &    27.0  &   3.1810  &   0.5963  &   1.9396  &   0.4694  &   1.2414  &   0.3615  &   0.0000  &   0.0678  \\
\hline
\hline
 X\# & Var Status & Data Set & $L_2^{\rm MOS2}$ & $cr_{\rm MOS2}$ & $\sigma_{\rm MOS2}$ & $cr_{\rm MOS2}^{\rm S}$ & $\sigma_{\rm MOS2}^{\rm S}$ & $cr_{\rm MOS2}^{\rm M}$ & $\sigma_{\rm MOS2}^{\rm M}$ & $cr_{\rm MOS2}^{\rm H}$ & $\sigma_{\rm MOS2}^{\rm H}$ \\
$[1]$ &  $[2]$  &  $[3]$  &  $[22]$ &   $[23]$  &  $[24]$   & $[25]$    & $[26]$    & $[27]$    &  $[28]$   &  $[29]$   &  $[30]$  \\[+1mm]
\hline
  1 &   Uncert  &     3   &   16.9  &   3.7830  &   0.9184  &   2.4310  &   0.7169  &   1.3520  &   0.5082  &   0.0000  &   0.2671 \\
    &           &     5   &   21.8  &   3.3101  &   0.7358  &   1.2515  &   0.5254  &   1.9346  &   0.4799  &   0.1240  &   0.1872 \\
  2 &   Const   &     3   &   16.9  &   3.7571  &   1.0486  &   0.0000  &   0.2600  &   1.8097  &   0.6987  &   1.9474  &   0.7375 \\
    &           &     5   &   19.2  &   3.6836  &   0.9986  &   0.0000  &   0.1867  &   0.8314  &   0.5721  &   2.8522  &   0.7969 \\
  3 &   Noinfo  &     3   &    5.3  &   1.2067  &   0.5056  &   0.0287  &   0.1578  &   0.3144  &   0.2606  &   0.8636  &   0.4035 \\
  4 &   Var     &     3   &   35.8  &   5.1319  &   0.9696  &   1.5404  &   0.6497  &   2.7681  &   0.6045  &   0.8234  &   0.3904 \\
    &           &     4   &   75.6  &   7.5895  &   1.0367  &   1.4724  &   0.5831  &   4.3611  &   0.7020  &   1.7560  &   0.4919 \\
    &           &     5   &   45.2  &   4.1332  &   0.7263  &   0.2420  &   0.3470  &   2.9420  &   0.5314  &   0.9493  &   0.3531 \\
  5 &   Uncert  &     3   &    1.9  &   0.6367  &   0.5123  &   0.6367  &   0.3823  &   0.0000  &   0.1562  &   0.0000  &   0.3032 \\
    &           &     4   &   34.9  &   3.3469  &   0.6648  &   1.6299  &   0.4648  &   1.6700  &   0.4481  &   0.0470  &   0.1584 \\
    &           &     5   &   14.5  &   1.7518  &   0.4607  &   1.0534  &   0.3500  &   0.6985  &   0.2908  &   0.0000  &   0.0721 \\
  6 &   Const   &     3   &    7.7  &   1.3382  &   0.6299  &   0.0000  &   0.3383  &   0.0900  &   0.2523  &   1.2483  &   0.4676 \\
    &           &     5   &   15.7  &   1.9042  &   0.5742  &   0.0000  &   0.2060  &   0.4021  &   0.3204  &   1.5021  &   0.4297 \\
  7 &   Const   &     3   &    3.6  &   1.7374  &   0.7574  &   0.6287  &   0.5433  &   0.9286  &   0.4724  &   0.1802  &   0.2353 \\
    &           &     4   &    1.7  &   0.6576  &   0.5783  &   0.0000  &   0.3803  &   0.6576  &   0.3924  &   0.0000  &   0.1893 \\
  8 &   Const   &     3   &   22.2  &   3.4013  &   0.7177  &   2.2301  &   0.5873  &   1.1712  &   0.4071  &   0.0000  &   0.0666 \\
    &           &     4   &   22.4  &   2.9989  &   0.6168  &   1.7462  &   0.4723  &   1.2527  &   0.3939  &   0.0000  &   0.0468 \\
    &           &     5   &   18.6  &   2.6419  &   0.5682  &   1.4887  &   0.4440  &   1.1532  &   0.3414  &   0.0000  &   0.0958 \\
\hline
\end{tabular}

\footnotesize
The first column gives the source number. The second column provides
the set-to-set variability status of the source (see section 5.2).
The third column gives the data set number in which the count rates
of the source were measured. Cols. 4-12 (resp. 13-21 and 22-30) give
the logarithmic likelihood $L_2$ for the given instrument, the total
count rate $cr$ in the whole energy band (0.4-10.0\,keV) and
its associated error ($\sigma$), the count rates in the three
different energy bands ($S:[0.4-1.0\,keV], M:[1.0-2.5\,keV],
H:[2.5-10.0\,keV]$) and their errors. The count rates and the related
uncertainties are all expressed in $10^{-3}$\,counts\,s$^{-1}$.

\end{table*}

Fluxes and luminosities can be calculated from count rates provided that the
response function of the telescope is known and certain assumptions are made
about the emission model, interstellar absorption and distance to the
region. We used {\tt xspec} to calculate the conversion factors. A
combination of models {\tt wabs*apec} was used. Here {\tt apec} is an
emission spectrum from a hot optically thin thermal plasma. The APEC code
(Smith et al., \cite{Smith01}) is a modern version of the well-known
model of Raymond \& Smith (\cite{RS}). {\tt wabs} is an interstellar
medium (ISM) absorption model. As we are most interested in X-ray
properties of hot massive stars, the plasma temperature in the {\tt apec}
model was set to $kT=0.6$\,keV, a value typical for X-ray emission produced
by shocks formed in stellar winds of early-type stars. The interstellar
absorption over the field is relatively uniform with
$E(B-V)=0.52$\,mag (Massey \& Johnson, \cite{MJ93}). Using a
well-known transformation from Bohlin et al. (\cite{Bohlin78})
$N_H=5.8\times{10^{21}}\times{E(B-V)}$ this translates to a column
density $N_H=3\times{10^{21}}\,{\rm cm^{-2}}$. We used this value in
the {\tt wabs} model. Using the above combination of models and {\sl
XMM-Newton} response function we derive the conversion factors for
{\sc mos} and {\sc pn} instruments in the 0.4\,--\,10.0\,keV 
energy range as

$$
F_{a,MOS} = 0.76\times 10^{-11}\rm \,ergs\,cm^{-2}\,s^{-1} \times cr
$$
$$
F_{a,PN} = 0.23\times 10^{-11}\rm \,ergs\,cm^{-2}\,s^{-1} \times cr
$$
$$
F_{0,MOS} = 2.01\times 10^{-11}\rm \,ergs\,cm^{-2}\,s^{-1} \times cr
$$
$$
F_{0,PN} = 0.61\times 10^{-11}\rm \,ergs\,cm^{-2}\,s^{-1} \times cr
$$

\noindent where $cr$ is the corresponding count rate, the
indices ``a'' and ``0'' mark the absorbed and unabsorbed fluxes
respectively. Luminosities can be readily obtained from the above fluxes
assuming a distance to the nebula of 2.5\,kpc (see section 6). We caution
that the conversion from count rates to energy is very model-dependent and
can only be used as a very rough estimation of the flux. Below we
provide individual fluxes for the brightest sources which allow spectral
fitting.

\subsection{The detection limit}

Evaluation of the detection limit of our observations is not a trivial
task. First, this limit is {\em a priori} not uniform throughout the field
of view as the {\sl XMM-Newton} effective exposure duration is decreasing
from the center of the FOV towards the edges. Second, as different sources
in our catalogue were detected in different observations or their
combinations, the detection limits in the intermediate partial source lists
are different.

\begin{figure}
\centering
\includegraphics[width=8.5truecm]{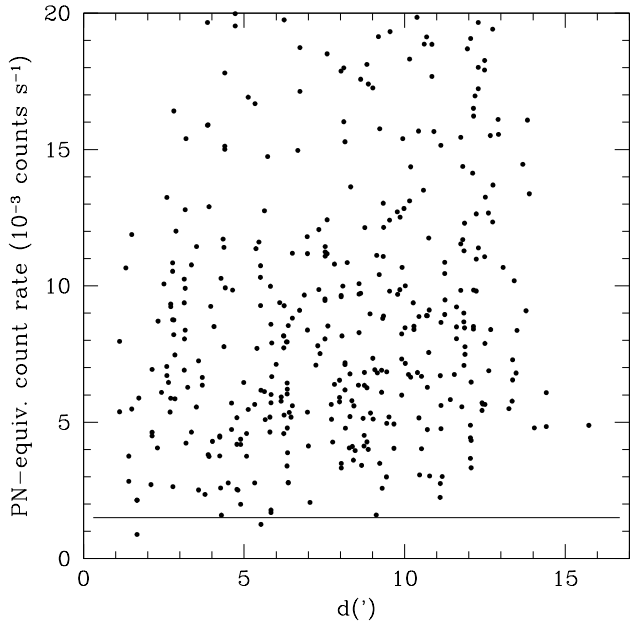}

\caption{Bottom part of the distribution of the {\sc pn}-equivalent
count rates of the X-ray sources as a function of their distance $d$ to the
center of the FOV. The horizontal line shows the adopted detection limit.}

\label{fig_detlimit}
\end{figure}

Thus, to estimate the overall limit, we adopted a completely
empirical approach, taking advantage of the large number of X-ray
sources in the field. We assumed that a good indication of the
detection limit in the different parts of the field is given by the
brightness of the faintest sources detected in these areas. Due to
the presence of gaps, not all sources which are detected in the {\sc
mos} images, are also present in the {\sc pn} ones. To increase
statistics, we first computed an equivalent {\sc pn} count rate, in
the range 0.4-10.0\,keV, for each source missing in the {\sc pn}
images. To first order, the relation between the count rates
measured in any of the two {\sc mos} detectors and in the {\sc pn}
detector is linear. We found the conversion factor using the count
rates of the sources, which were detected in all three {\sc epic}
instruments. Figure\,\ref{fig_detlimit} displays the source {\sc
pn}-equivalent count rates as a function of the distance from the
central axis of the FOV. A lower limit is clearly seen in the
distribution. Towards the edges of the FOV, the limit is increased
as expected. The faintest sources in the region have the {\sc
pn}-equivalent count rate about $1.5 \times
10^{-3}$\,counts\,s$^{-1}$, which we accept as our detection limit.
In terms of flux, using the conversion above, this limit amounts to
$3.3 \times 10^{-15}\rm \,erg\,cm^{-2}\,s^{-1}$. 

\subsection{Cross-identification}

\begin{figure}
\begin{center}
\includegraphics[width=8.5truecm]{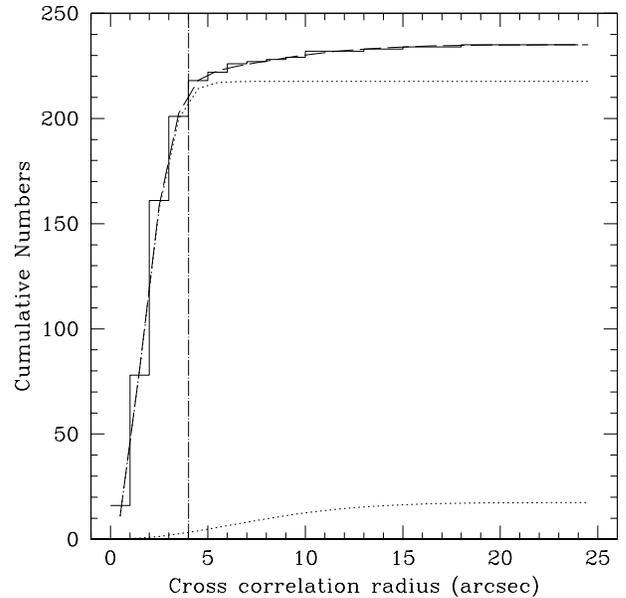}
\end{center}

\caption{Cumulative numbers of correlations between the X-ray detections and
the 2MASS catalogue objects as a function of the correlation radius. The
dotted curves correspond to the best fitting expressions for the real and
spurious correlations. The dashed curve yields the sum of these terms and
the dash-dotted vertical line corresponds to the adopted correlation radius
of 4.0\arcsec.}

\label{fig_id}
\end{figure}

We have cross-correlated the positions of the 235 sources with various
optical and infrared catalogues. The 2MASS catalogue provides the most
complete coverage of the Carina complex. We therefore selected this
catalogue in order to determine the optimal radius of cross-correlation. To
this aim, we adopted the approach outlined by Jeffries et al.\
(\cite{Jeff97}) and applied to the {\sl XMM-Newton} data of the young open
clusters NGC\,6530 and NGC\,6383 by Rauw et al.\,(\cite{Rauw02},
\cite{Rauw03}). The distribution of the cumulative number of catalogued
sources as a function of the cross-correlation radius
$r$ is given by

\begin{eqnarray*}
\Phi(d \leq r) & = & A\,\left[1 - \exp{\left(\frac{-r^2}{2\,\sigma^2}\right)}\right] \\
& + & (N - A)\,\left[1 - \exp(-\pi\,B\,r^2)\right]
\end{eqnarray*}

\noindent where $N$, $A$, $\sigma$ and $B$ stand respectively for
the total number of cross-correlated X-ray sources ($N = 235$), the number
of true correlations, the uncertainty on the X-ray source position and the
surface density of catalogue sources. For further details on the method we
refer to the work of Jeffries et al.\ (\cite{Jeff97}). The fitting
parameters $A$, $\sigma$ and
$B$ were obtained from the best fit to the actual distribution
displayed in Fig.\,\ref{fig_id}. For the 2MASS catalogue, we derive
$A=217.6$, $\sigma = 1.6$\arcsec and $B=0.004$\,arcsec$^{-2}$.
The optimal radius that includes the majority of the true
correlations while simultaneously limiting contamination by spurious
correlations, is found to be around 4.0\arcsec. We thus
consider an infrared source as a possible counterpart if it falls
within 4.0\arcsec of the coordinates of the X-ray source. This
is significantly smaller than the corresponding optimal radius for
the {\sl XMM} data on NGC\,6383. The main reason is the larger
surface density of 2MASS sources in the Carina region. For $r =
4.0\arcsec$, we expect statistically to achieve 208 true and
6 spurious correlations.

We have also cross-correlated the positions of our X-ray sources with
the AC03 {\sl XMM-Newton} and EV03 {\em Chandra} X-ray surveys of the field,
using a similar technique. The optimal cross-correlation radii were found to
be around 5.0\arcsec for the AC03 and 4.5\arcsec for the EV03 data
statistically providing around 58/3 and 54/5 true/false correlations
respectively.

The results of the correlations with the 2MASS and X-ray catalogues
and the results of various optical studies of the Carina Nebula are listed
in Table\,\ref{tbl-ident}. The source designation in this table follows the
naming conventions recommended by the IAU for serendipitous sources detected
with {\sl XMM-Newton}: the XMMU\,J prefix is followed by the right ascension
HHMMSS.s (in hours, minutes, seconds and tenths of seconds, equinox J2000)
and the declination of the source +/$-$DDMMSS (in degrees, arcminutes and
arcseconds, equinox J2000), both truncated not rounded. The
designations of the sources identified with the AC03 catalogue are taken
from that catalogue; note that we still provide our own coordinates of such
sources. 167 EPIC sources have a single 2MASS optical counterpart
within a radius of less than 4.0\arcsec. The average angular
separation between the X-ray source and the optical counterpart is
$(2.3 \pm 0.9)$\arcsec. Cudworth et al.\ (\cite{Cud93}) performed a proper
motion study of the open clusters in the Carina complex. Using photographic
plates spanning about a century, they derived membership probabilities for
577 stars. We have cross-correlated the astrometry of our EPIC sources with
the catalogue of Cudworth et al. Information on proper motion membership
probabilities is provided in the last column of Table\,\ref{tbl-ident}.

\begin{sidewaystable*}

\caption{Sample of the cross identification table of X-ray sources
detected in our EPIC images of the Carina Nebula.}

\tiny

\label{tbl-ident}
\begin{tabular}{r c c c c c c c r r r r r r l l c r}
\hline
X\#& XMMU\,J & RA & DEC & AC03 & EV03 & \multicolumn{5}{c}{2MASS}                   & \multicolumn{6}{c}{Optical}\\
   &         &    &     &      &      & Nr. & $d$ &\multicolumn{1}{c}{$J$} & \multicolumn{1}{c}{$H$} & \multicolumn{1}{c}{$K_S$} & \multicolumn{1}{c}{$V$}        & \multicolumn{1}{c}{$B-V$} & \multicolumn{1}{c}{$U-B$} & \multicolumn{1}{c}{Name} &  \multicolumn{1}{c}{Spectral} & Ref. & \multicolumn{1}{c}{Member}\\
   &         &    &     &      &      &     & (\arcsec)   & & &       &       &       &       &      &  \multicolumn{1}{c}{Type}   &      & \multicolumn{1}{c}{(\%)} \\
\multicolumn{1}{c}{$[1]$} & $[2]$ & $[3]$ & $[4]$ & \multicolumn{1}{c}{$[5]$} &  \multicolumn{1}{c}{$[6]$} &  \multicolumn{1}{c}{$[7]$} & $[8]$ & \multicolumn{1}{c}{$[9]$} & \multicolumn{1}{c}{$[10]$} & \multicolumn{1}{c}{$[11]$} & \multicolumn{1}{c}{$[12]$} & \multicolumn{1}{c}{$[13]$} & \multicolumn{1}{c}{$[14]$} & \multicolumn{1}{c}{$[15]$} & \multicolumn{1}{c}{$[16]$} & \multicolumn{1}{c}{$[17]$} & \multicolumn{1}{c}{$[18]$} \\
\hline
   1 &  104236.8-594357  &  10 42 36.83  &  -59 43 57.9  &   -   &   -  & 0 & 3.9 & 13.00 & 12.37 & 12.18 &       &      &         &                                     &                        &       &    \\
   2 &  104256.6-594220  &  10 42 56.61  &  -59 42 20.2  &   -   &   -  & 2 &     &       &       &       &       &      &         &                                     &                        &       &    \\
   3 &  104303.2-593949  &  10 43 03.21  &  -59 39 49.3  &   -   &   -  & 0 &     &       &       &       &       &      &         &                                     &                        &       &    \\
   4 &  104304.4-594902  &  10 43 04.43  &  -59 49 02.7  &   -   &   -  & 1 & 3.9 & 16.15 & 14.41 & 14.20 & 12.47 & 0.38 & $-0.80$ & MJ 3                                &                        & MJ    &    \\
   5 &  104307.8-593557  &  10 43 07.82  &  -59 35 57.4  &   -   &   -  & 1 & 3.7 & 13.76 & 13.05 & 12.83 &       &      &         &                                     &                        &       &    \\
   6 &  104310.5-595012  &  10 43 10.56  &  -59 50 12.6  &   -   &   -  & 0 &     &       &       &       &       &      &         &                                     &                        &       &    \\
   7 &  104310.9-595124  &  10 43 10.91  &  -59 51 24.7  &   -   &   -  & 1 & 1.4 & 13.75 & 12.97 & 12.57 &       &      &         &                                     &                        &       &    \\
   8 &  104311.4-594423  &  10 43 11.40  &  -59 44 24.0  &   -   &   -  & 1 & 3.3 &  8.70 &  8.62 &  8.50 &  9.64 & 0.26 & $-0.76$ & HDE\,303316                         &                        & MJ    &    \\
   9 &  104315.1-593344  &  10 43 15.12  &  -59 33 44.2  &   -   &   -  & 1 & 3.1 & 12.77 & 12.30 & 12.14 &       &      &         &                                     &                        &       &    \\
  10 &  104317.5-593321  &  10 43 17.55  &  -59 33 21.7  &   -   &   -  & 0 &     &       &       &       &       &      &         &                                     &                        &       &    \\
  11 &  104319.8-594248  &  10 43 19.86  &  -59 42 48.5  &   -   &   -  & 1 & 2.9 & 15.15 & 14.15 & 13.58 &       &      &         &                                     &                        &       &    \\
  12 &  104320.6-594619  &  10 43 20.63  &  -59 46 19.8  &   -   &   -  & 1 & 2.5 & 14.45 & 13.52 & 13.21 &       &      &         &                                     &                        &       &    \\
  13 &  104325.9-594720  &  10 43 25.93  &  -59 47 20.5  &   -   &   -  & 0 &     &       &       &       &       &      &         &                                     &                        &       &    \\
  14 &  104326.8-595251  &  10 43 26.90  &  -59 52 51.5  &   -   &   -  & 0 &     &       &       &       &       &      &         &                                     &                        &       &    \\
  15 &  104327.4-594202  &  10 43 27.48  &  -59 42 02.4  &   -   &   -  & 0 &     &       &       &       &       &      &         &                                     &                        &       &    \\
  16 &  104329.8-593344  &  10 43 30.03  &  -59 33 48.0  &   1   &   -  & 1 & 3.0 & 13.41 & 12.52 & 12.21 &       &      &         &                                     &                        &       &    \\
  17 &  104333.7-594030  &  10 43 33.76  &  -59 40 30.0  &   -   &   -  & 1 & 2.3 & 15.49 & 14.69 & 14.49 &       &      &         &                                     &                        &       &    \\
  18 &  104339.9-593445  &  10 43 39.92  &  -59 34 50.9  &   -   &   -  & 1 & 2.3 & 13.21 & 12.26 & 11.96 & 15.43 & 1.35 &         & DETWC-14 18                         &                        & DETWC &    \\
  19 &  104340.7-594638  &  10 43 40.75  &  -59 46 38.4  &   -   &   -  & 0 &     &       &       &       &       &      &         &                                     &                        &       &    \\
  20 &  104341.2-593355  &  10 43 41.23  &  -59 33 55.7  &   -   &   -  & 1 & 3.3 & 11.23 & 10.80 & 10.55 & 17.82 & 0.75 &         & DETWC-14 29                         &                        & DETWC &    \\
  21 &  104341.3-593205  &  10 43 41.49  &  -59 32 09.8  &   3   &   -  & 1 & 3.5 & 12.70 & 11.97 & 11.71 & 15.72 & 1.30 & 0.69    & DETWC-14 43                         &                        & DETWC &    \\
  22 &  104341.4-594223  &  10 43 41.61  &  -59 42 26.8  &   4   &   -  & 2 &     &       &       &       &       &      &         &                                     &                        &       &    \\
  23 &  104343.0-594411  &  10 43 43.04  &  -59 44 11.3  &   -   &   -  & 1 & 3.4 & 15.63 & 14.31 & 13.27 &       &      &         &                                     &                        &       &    \\
  24 &  104343.1-593555  &  10 43 43.16  &  -59 35 55.7  &   -   &   -  & 1 & 3.9 & 14.41 & 13.49 & 13.12 &       &      &         &                                     &                        &       &    \\
  25 &  104344.6-593459  &  10 43 44.64  &  -59 34 59.1  &   -   &   -  & 2 &     &       &       &       &       &      &         & DETWC-14 76, 65                     &                        &       &    \\
  26 &  104345.0-595327  &  10 43 45.08  &  -59 53 27.7  &   -   &   -  & 1 & 2.3 & 10.02 &  9.99 &  9.97 & 10.43 & 0.14 & $-0.77$ & MJ 126                              &                        & MJ    &    \\
  27 &  104345.4-594159  &  10 43 45.43  &  -59 41 59.8  &   -   &   -  & 1 & 3.3 & 14.37 & 12.99 & 12.18 &       &      &         &                                     &                        &       &    \\
  28 &  104345.6-593431  &  10 43 45.60  &  -59 34 31.6  &   -   &   -  & 1 & 2.7 & 13.24 & 12.42 & 12.17 &       &      &         & DETWC-14 78, 81                     &                        &       &    \\
  29 &  104346.0-593221  &  10 43 46.05  &  -59 32 21.6  &   -   &   -  & 1 & 0.1 & 15.26 & 14.12 & 13.51 &       &      &         &                                     &                        &       &    \\
  30 &  104346.3-593256  &  10 43 46.36  &  -59 32 56.6  &   -   &   -  & 1 & 3.2 &  8.66 &  8.52 &  8.50 &  9.65 & 0.20 & $-0.63$ & CPD$-58^{\circ}$\,2611 = Tr\,14 20  & O6\,V((f))             & MJ    & 96 \\
  31 &  104346.3-593116  &  10 43 46.39  &  -59 31 16.5  &   -   &   -  & 1 & 2.6 & 12.37 & 11.70 & 11.51 &       &      &         &                                     &                        &       &    \\
  32 &  104347.3-593244  &  10 43 47.35  &  -59 32 44.2  &   -   &   -  & 1 & 3.0 & 12.05 & 11.56 & 11.30 & 15.20 & 1.36 & 0.77    & DETWC-14 107                        &                        & DETWC &    \\
  33 &  104348.8-593324  &  10 43 48.88  &  -59 33 24.6  &   -   &   -  & 1 & 1.4 &  9.55 &  9.39 &  9.26 & 10.73 & 0.35 & $-0.55$ & Tr\,14 21                           & O9\,V                  & MJ    & 81 \\
  34 &  104349.2-593404  &  10 43 49.22  &  -59 34 05.0  &   -   &   -  & 1 & 2.0 & 14.30 & 13.13 & 12.54 &       &      &         &                                     &                        &       &    \\
  35 &  104349.8-594453  &  10 43 49.45  &  -59 44 56.9  &   8   &   -  & 1 & 2.0 & 12.67 & 12.20 & 12.06 & 14.34 & 0.77 &         & Tr\,14 Y 334                        &                        & Cud   & 0  \\
  36 &  104351.0-595239  &  10 43 51.03  &  -59 52 40.0  &   -   &   -  & 1 & 2.9 & 12.00 & 11.60 & 11.50 & 13.75 & 0.41 & $-0.09$ & MJ 156                              &                        & MJ    &    \\
  37 &  104351.4-593117  &  10 43 51.47  &  -59 31 17.1  &   -   &   -  & 1 & 2.4 & 13.56 & 12.89 & 12.63 &       &      &         &                                     &                        &       &    \\
  38 &  104351.6-593244  &  10 43 51.63  &  -59 32 44.3  &   -   &   -  & 1 & 3.1 & 13.00 & 12.27 & 11.94 & 16.38 & 1.26 &         & DETWC-14 160                        &                        & DETWC &    \\
  39 &  104352.1-593556  &  10 43 52.02  &  -59 35 58.0  &   9   &   -  & 1 & 1.9 & 13.28 & 12.55 & 12.22 & 16.58 & 1.44 &         & DETWC-14 167                        &                        & DETWC &    \\
  40 &  104352.3-593924  &  10 43 52.48  &  -59 39 24.7  &  10   &   -  & 1 & 2.8 & 13.03 & 11.91 & 11.20 &       &      &         &                                     &                        &       &    \\
  41 &  104355.0-593130  &  10 43 55.09  &  -59 31 30.0  &   -   &   -  & 1 & 1.7 & 13.26 & 12.51 & 12.26 & 16.53 & 1.46 &         & DETWC-14 200                        &                        &       &    \\
  42 &  104355.2-593200  &  10 43 55.23  &  -59 32 00.7  &   -   &   -  & 1 & 2.4 & 10.51 & 10.04 &  9.70 & 12.52 & 0.78 & 0.43    & Tr\,14 14                           &                        & DETWC & 78 \\
  43 &  104355.9-594926  &  10 43 55.92  &  -59 49 26.1  &   -   &   -  & 0 &     &       &       &       &       &      &         &                                     &                        &       &    \\
  44 &  104356.0-593515  &  10 43 56.03  &  -59 35 15.1  &   -   &   -  & 0 &     &       &       &       &       &      &         &                                     &                        &       &    \\
  45 &  104356.7-593252  &  10 43 57.39  &  -59 32 54.8  & 11c   & 103  & 2 &     &       &       &       &       &      &         & HD\,93129 A + B                     &                        &       &    \\
\hline
\end{tabular}

\footnotesize
The first and second columns yield the number of the X-ray source as
well as the name according to the naming conventions for
serendipitous {\it XMM-Newton} sources. To avoid duplicating the
nomenclature, AC03 names are given for the sources cross-identified
with the AC03 catalogue. The third and fourth columns provide the
{\sl XMM-Netwon} coordinates of the sources. Columns $[5]$ and
$[6]$ list the star number in the X-ray catalogues of AC03 and EV03
according to our identification. Columns $[7]$ to $[11]$ summarize
the results of the cross-correlation with the 2MASS catalogue. The
columns labelled `Nr.' and `$d$' yield the number of 2MASS
counterparts within a 4.0\arcsec radius and the angular separation
between the X-ray source and the IR counterpart. Columns $[12]$ to
$[17]$ provide information on the properties of optical counterparts
(when available). DETWC-14 and DETWC-16 names yield the sequence
number of the optical star in the catalogue of DeGioia-Eastwood et
al.\ (\cite{DeG01}) for the Tr\,14 and Tr\,16 clusters respectively.
In a similar way, Tr\,14 and Tr\,16 numbers refer to the numbering
scheme introduced by Feinstein et al. (\cite{Feinst73}) \ whilst
Tr\,14 Y and Tr\,16 Y correspond to the convention of Cudworth et
al.\ (\cite{Cud93}). Finally, MJ numbers are taken from Massey \&
Johnson (\cite{MJ93}). The column labelled `Ref' yields the reference
for the optical photometry whilst the last column indicates the
membership probability from the proper motion study of Cudworth et
al.\ (\cite{Cud93}).

\end{sidewaystable*}

Finally, given the limiting sensitivity of $1.5 \times
10^{-3}$\,counts\,s$^{-1}$ for {\sc pn} and the size of the field of view
(roughly 0.25 square degree), we can use the $\log{N}$ -- $\log{S}$ relation
of Motch et al.\ (\cite{Motch03}) for $b \sim 0^{\circ}$ to estimate the
number of field stellar sources. In this way, we find that about 50 -- 70
field stars could be detected. Since the Motch et al.\ (\cite{Motch03})
relation was established from fields that do not harbour star formation
regions, the significantly larger number of sources detected in our Carina
field clearly results from X-ray sources associated with the clusters in the
field of view.

\subsection{Extragalactic background sources}

The line of sight towards $\eta$\,Car ($l_{\rm II} = 287.60^{\circ}$,
$b_{\rm II} = -0.63^{\circ}$) is nearly tangent to the Carina spiral arm. As
a result, the neutral hydrogen column density along this direction must be
quite large and should produce a substantial absorption of X-ray photons
from extragalactic background sources (EBS).

To get a rough first order estimate of the total Galactic extinction along
this line of sight, we made use of the {\it DIRBE/IRAS} extinction maps
provided by Schlegel et al.\ (\cite{Schlegel98}). Schlegel et al.\ caution
that one has to be careful when using these maps near the Galactic plane.
Moreover, the interstellar extinction in the Carina region is rather patchy
and its properties could deviate from those of the average Galactic
extinction curve (e.g.\ Carraro et al.\ \cite{Carraro04}). With these
limitations in mind, we find that the {\it DIRBE/IRAS} maps indicate a
rather large, but highly variable $E(B-V)$ colour excess (between $\sim 6.4$
on average at the edges of the EPIC field and $\sim 14.5$ near $\eta$\,Car).
Converting these values into neutral hydrogen column densities by means of
the gas to dust ratio of Bohlin et al.\ (\cite{Bohlin78}), we estimate
$N_{\rm H}$ in the range $3.7$ -- $8.5
\times 10^{22}$\,cm$^{-2}$.

Assuming that EBSs (most of which are probably Active Galactic Nuclei
(AGN)) have a power-law spectrum with a photon index of 1.4 (Giacconi
et al.\ \cite{Giacconi01}), and suffer a total interstellar absorption of
$3.7$ -- $8.5 \times 10^{22}$\,cm$^{-2}$, the above detection limits
translate into unabsorbed fluxes of $1.0$ -- $1.5 \times
10^{-14}$\,erg\,cm$^{-2}$\,s$^{-1}$ and $2.8$ --
$4.3 \times 10^{-14}$\,erg\,cm$^{-2}$\,s$^{-1}$ in the 0.5 -- 2.0\,keV and
2.0 -- 10\,keV band respectively. From the $\log{N}$ -- $\log{S}$ relation
of Giacconi et al.\ (\cite{Giacconi01}), we expect to detect between 
$\sim 16$ and 24 extragalactic background sources.

To discriminate these sources from the Galactic ones, the following
basic properties of AGNs could in principle be used (see e.g. Fiore
\cite{Fiore97}, Nandra \cite{Nandra01}, Bauer et al. \cite{Bauer04}):
\begin{enumerate}
\item The spectra of AGNs are power laws with the canonical intrinsic
X-ray spectral slope $\Gamma\sim 1.9-2.0$. The average observed spectral
slope of EBS is about $1.4$ (Giacconi et al. \cite{Giacconi01}).
\item Their flux is variable on timescales from a few hours to years.
\end{enumerate}

In practice, it is impossible to fit the individual source spectra
since the candidates are very faint and the count statistics poor.
For this reason, we have simulated the values of two {\sc pn}
hardness ratios $HR_1=(M-S)/(M+S)$ and $HR_2=(H-M)/(H+M)$ for a grid
of simple models. Within {\tt xspec}, we have used the {\sc pn}
response matrices to generate fake spectra corresponding to
differently absorbed power law spectra. We considered three photon
indices $\Gamma = 1.4, 1.7, 2.0$ and several neutral hydrogen
densities $N_{\rm H} = (0.1, 0.3, 1.0, 3.0, 7.0, 11.0, 15.0) \times
10^{22}$\,cm$^{-2}$. We then compared these simulated hardness
ratios with the observed ones of all discrete sources not
identified with optical or infra-red catalogues. For the list of the
potential EBS candidates, we select those sources whose errorboxes
cover the simulated hardness ratios. The resulting EBS
candidates are the sources 3, 6, 119, 124, 170, 185, 207. Their
total number is seven which is lower than the estimation above;
still the estimation gives only a rough idea about the number of the
background sources, given the uncertainties involved.

\section{Open clusters in the Carina Nebula\label{sect_car_review}}

The region covered by our EPIC data harbours several young open clusters
that are extremely rich in hot massive stars; the most important ones being
Trumpler\,14 and 16. Tr\,14 is a compact cluster containing three very hot
O3 stars. Tr\,14 is probably emerging from the eastern and near side of the
dense parental molecular cloud (Tapia et al.\ \cite{Tapia03}). Tr\,14 might
be slightly younger than Tr\,16 as suggested by a fainter absolute magnitude
and a larger He\,{\sc ii} $\lambda$\,4686/ He\,{\sc ii} $\lambda$\,4541
ratio in the spectra of early and mid O-type dwarfs (Walborn \cite{Walb95}).
Collinder\,232 is probably not a physical entity but a condensation of stars
in the vicinity of Tr\,14 and 16 (see e.g.\ Walborn \cite{Walb95} and Tapia
et al.\ \cite{Tapia03}). In a similar way, Collinder\,228 could actually be
part of Tr\,16. The apparent distinction between the two clusters might
result from obscuration by a dust lane (Walborn \cite{Walb95}). Note an
alternative view of Carraro et al. (\cite{Carraro04}) who argue that
Collinder\,232 may be a physical aggregate (a rather sparse young open
cluster), based on the comparison of its colour-magnitude diagrams and
theoretical zero age main sequence (ZAMS) tracks.

The extinction in the Carina complex has been subject to much controversy.
Conflicting conclusions ranging from a normal uniformly high to an anomalous
and variable extinction have been proposed. This dilemma has important
consequences for a consistent determination of the distance moduli (see
e.g.\ the discussion in Walborn \cite{Walb95}). In addition, since the line
of sight is along the direction of a galactic spiral arm, there are probably
many foreground and background objects that appear projected onto the Carina
Nebula.

The exact ages of the clusters and the star formation history of the
Carina Nebula are other tricky issues. Up until recent years,
evidence for ongoing star formation in the Carina Nebula was rather
scarce. Megeath et al.\ (\cite{Meg96}) and Rathborne et al.\
(\cite{Rath02}) reported evidence of star formation triggered in the
molecular cloud exposed to the ionizing radiation of the hot stars
of the Carina Nebula. None of the three near-IR sources N\,1,
N\,3 and N\,4 proposed to be associated with embedded recently born
O-stars is detected in our X-ray data. More recently, Smith et al.\
(\cite{Smith03}) reported the discovery of dozens of candidates of
so-called proplyds in the Carina Nebula, clearly demonstrating that
the formation of low and intermediate mass stars is actively
proceeding. We note that none of the objects discussed by Smith et
al.\ (\cite{Smith03}) is detected as an X-ray source in our EPIC
data.

DeGioia-Eastwood et al.\ (\cite{DeG01}) obtained
$UBV$ photometry of the Carina clusters. They found a significant population
of pre-main sequence stars in Tr\,14 and Tr\,16 that have been forming over
the last 10\,Myr whilst the most massive stars in Tr\,16 have ages between 1
and 3\,Myr (with one exception, a star in the 5-6\,Myr range in each
cluster). According to their results, intermediate mass star formation seem
to have proceeded continuously over the past 10\,Myr and the formation of
intermediate mass stars started well before that of OB stars and was
apparently not disrupted by the formation of these OB stars. Finally,
Tapia et al.\ (\cite{Tapia03}) performed an extended $UBVRIJHK$ photometric
study of the clusters in the Carina Nebula. They infer ages between less
than 1\,Myr and 6\,Myr for Tr\,14 and 16. A small number of IR excess stars
were found in both clusters. Tapia et al.\ further detected a
population of 19 near-IR sources in the Car I H{\sc ii} region whose
formation could have been triggered by the action of the early-type stars.
We note that none of these objects is detected in our EPIC data.

Corcoran (\cite{Corc99}) analysed ROSAT-HRI observations of the Carina
complex. While he found the $L_{\rm X}$/L$_{\rm bol}$ ratio of early-type
stars to spread between $10^{-8}$ and $10^{-6}$, he did not find evidence
for a population of X-ray bright pre-main sequence (PMS) stars. However, he
cautioned that the failure to detect these objects could simply be due to
the limited sensitivity of the ROSAT satellite.\\

Trumpler\,16 is particularly rich in massive binaries (see e.g.\
Levato et al.\ \cite{Lev91}). The properties of several eclipsing or
ellipsoidal early-type binaries (HD\,93205, Morrell et al.\
\cite{Morr01}, Antokhina et al.\ \cite{Ant00}; Tr\,16-1, Freyhammer
et al.\ \cite{Frey01}; and Tr\,16-104, Rauw et al.\ \cite{Rauw01})
have been investigated recently. The components of the three
binaries are all found to lie pretty close to the zero-age main
sequence suggesting that they may be younger than 1\,Myr (Rauw
\cite{Rauw04}). The absolute magnitudes determined from the
study of these eclipsing binaries suggest a distance modulus (DM) of
$11.95 \pm 0.06$ (corresponding to a distance of 2.45\,kpc) lower
than the value of DM = $12.55 \pm 0.1$ (which corresponds to a
distance of 3.24\,kpc) derived by Massey \& Johnson (\cite{MJ93})
and Massey, DeGioia-Eastwood, \& Waterhouse (\cite{MGW01}) from a
spectoscopic and photometric study of the Trumpler 14 and 16
clusters. Allen \& Hillier (\cite{AH93}), Meaburn (\cite{Mea99}),
and Davidson et al. (\cite{Dav01}) investigated the geometry and
kinematics of the Homunculus Nebula and found the distance to
$\eta$\,Car around 2.3$\pm\sim$0.3\,kpc. Tapia et al.
(\cite{Tapia03}) derived the average distance to Tr\,14, Tr\,15 and
Tr\,16 to be $d=2.7$\,kpc from their {\sl UBVRIJHK} photometric
study, although they note a very large scatter in $d$. Carraro et
al. (\cite{Carraro04}) confirm the distance of 2.5\,kpc to Tr\,14
but reach a different conclusion about the distance to Tr\,16.
Comparing the colour-magnitude diagrams and the Hertzsprung-Russel
diagram (based on their {\sl UBVRI} photometry) with post and
pre-main sequence tracks and isochrones, they find that Tr\,16 lies
at the distance of about 4\,kpc from the Sun. EV03 adopted the
distance of 2.5\,kpc to the clusters in the Carina Nebula. In the
current study we also adopt the distance of 2.5\,kpc to all stars in
the region. Note that while the absolute value of the X-ray and
bolometric luminosities will change with the distance, their ratio
remains unchanged.

\begin{figure}
\begin{center}
\includegraphics[width=8.5truecm]{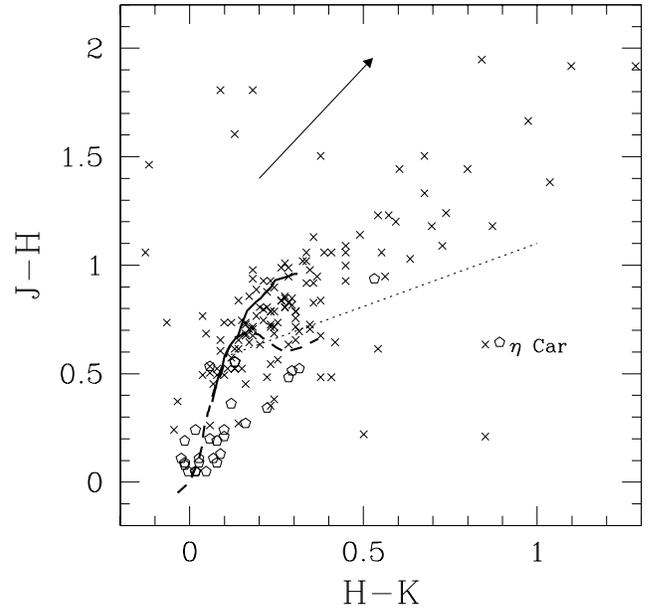}
\end{center}

\caption{$JHK_s$ colour-colour diagram of the single IR counterparts falling
within the 4.0\arcsec correlation radius from an X-ray source. The heavy
dashed and solid lines yield the intrinsic colours of main sequence
and giant stars respectively, whereas the reddening vector is
illustrated for $A_V = 5$ magnitudes of visual extinction. Bright
counterparts with $J < 10.0$ are displayed by open pentagons. The crosses
indicate fainter IR sources. The dotted straight line yields the locus of
dereddened colours of classical T\,Tauri stars according to Meyer et al.\
(\cite{Meyer97}).\label{2mass}}

\end{figure}

Fig.\,\ref{2mass} illustrates the $JHK_s$ colour-colour diagram of the
2MASS infrared counterparts identified in our current study. We have
included only those objects that were detected in all three filters
$J$\,(1.25\,$\mu$m), $H$\,(1.65\,$\mu$m) and
$K_s$\,(2.17$\mu$\,m) with a photometric accuracy better than $\sigma \leq
0.10$ and which are single counterparts of EPIC sources within a
correlation radius of 4.0\arcsec. The intrinsic colours of dwarfs and giants
(taken from Bessell \& Brett \cite{Bess88}) are indicated by the dashed
and solid lines respectively. The open pentagons correspond to stars
with $J$ magnitudes brighter than 10.0 (mostly moderately reddened
early-type stars), whilst the crosses indicate fainter stars.

In principle, this diagram can be used to identify pre-main sequence stars
(Lada \& Adams \cite{Lada92}) either through a large circumstellar extinction
characteristic of deeply embedded protostars or through infrared excesses
produced by circumstellar disks in classical T\,Tauri stars (cTTs). The
latter objects are expected to fall mostly to the right of the reddening
band. Meyer et al.\ (\cite{Meyer97}) have shown that the locus of dereddened
$JHK$ colours of cTTs can be described by a simple linear relation between
$J - H$ and $H - K$. This relation is shown in Fig.\,\ref{2mass} by the
dotted straight line. Only a modest number of the IR counterparts in
Fig.\,\ref{2mass} have colours that are consistent with these objects being
surrounded by large amounts of circumstellar material. The majority of
the secondary X-ray sources could thus be weak line T\,Tauri stars (wTTs)
which display little or no evidence for circumstellar material in their
spectra and photometry. This situation is strongly reminiscent of other
young clusters such as NGC\,6383 (Rauw et al. \cite{Rauw03}) or NGC\,6231
(Sana et al. \cite{Sana07}) where the X-ray selected pre-main sequence
objects are also dominated by wTTs.

\section{Variability}

\subsection{Short term variability}

To study possible variations of the detected sources within individual data
sets, we applied the following variability tests: (i)\,we computed $\chi^2$
to test the null hypothesis of a constant count rate level in the data;
(ii)\,we applied the Kolmogorov-Smirnov (KS) test to check if the
statistics of the count rate follows a normal distribution; (iii)\,we used
the so-called {\em probability of variability} {\em pov} test suggested
by Preibisch \& Zinnecker (\cite{Preib02}) in its modified version suggested
by Sana et al. (\cite{Sana04}).

In all methods, the source data were extracted from circular regions
with the radius equal to half of the distance to the nearest other
source, or 60\arcsec if no nearby sources were present. The first
two methods were applied to the background-corrected light curves of
every source produced by binning photons in time and energy (see
below). Note that, for faint sources, the KS test can sometimes fail
for the background-subtracted light curves. The third method was
applied on the series formed by photon arrival times (the photons
were still binned in the same energy bins as for the $\chi^2$ test).
Thus, when using the {\em pov} method, it was not possible to
account for the background. All variability tests were also applied
to the background data.

\begin{figure*}
\centering
\includegraphics[width=15.0truecm]{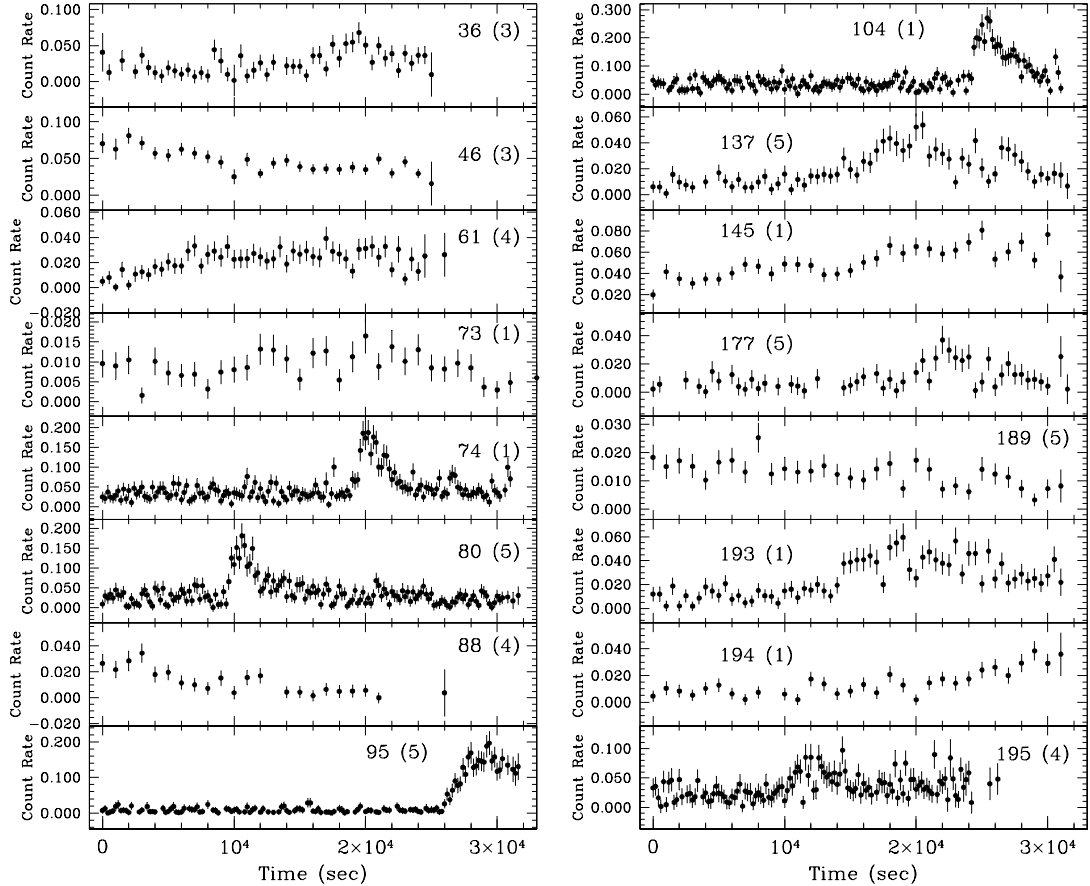}

\caption{{\sc pn} background-subtracted light curves in counts\,s$^{-1}$ of sources
showing significant short term variability ({\sc mos1} light curve for the source
73). The energy range is 0.4--10.0\,keV. The source numbers followed by the
data set numbers in parantheses are shown (for source identification see
text and Table\,\ref{tbl-ident}). In this and the following light
curve plots (Fig.\ref{fig-lc_074}--\ref{fig-lc_104}) the time shown is
relative to the start of the exposure.}

\label{fig-lc_shortvar}
\end{figure*}

To use the $\chi^2$ and KS criteria, we produced light curves of the
sources with various time bin sizes (from 10 to 5000 sec) and in the
soft $S$, medium $M$, hard $H$ and total (0.4--10.0\,keV) energy
ranges.  Since the Carina region is quite crowded, it was often
impossible to find a source-free area in the vicinity of a given
source to extract a background light curve. For this reason, we
adopted the following scheme when extracting the background light
curves. For the {\sc mos} instruments, the internal instrumental
background is more or less constant within a single CCD. For the
{\sc pn}, the internal background primarily varies with the distance
along the detector ``y''-axis from the central detector ``x''-axis
(separating 2 sets of the {\sc pn} CCDs). Thus, for the {\sc mos}
instruments, we defined several object-free circular areas on every
CCD and extracted one background light curve per energy range per
{\sc mos} CCD from all these areas. The background light curves from
a given {\sc mos} CCD were used for all objects located on that CCD.
We performed a similar procedure with the {\sc pn} background light
curves except that circular background areas were selected in 10
strips parallel to the central {\sc pn} ``detx''-axis. Apart from
instrumental background, there exists diffuse X-ray emission from
the Carina nebula. Thus the background light curves obtained as
above, may overestimate or underestimate the real background level
near particular sources. This could influence the results for the
faintest sources. However, in practice this is not very important as
(i)\,we use several methods for detecting variability, including
those without background subtraction; (ii)\,the count rates of the
faintest sources are small anyway resulting in large errors and
small chances to find variability; (iii)\,as the diffuse emission
has large spatial scale (about the size of one CCD or larger) the
adopted procedure roughly accounts for global changes of the diffuse
background.

To increase signal, the sum of the background-corrected {\sc mos1} and {\sc mos2}
light curves was also analysed whenever possible.
In Fig.\ref{fig-lc_shortvar} we show the {\sc pn} light curves of those objects
for which the null hypothesis is rejected at a confidence level of 99\%
using all three variability test methods (for source 81 we show the {\sc mos1}
light curve as the source falls onto a gap between {\sc pn} CCDs). We do not show
the {\sc mos1} and {\sc mos2} light curves due to lack of space. The behaviour of these
light curves is similar to the {\sc pn} ones. The time bin sizes of the plots were
selected according to typical variability time scales of the corresponding
light curves.

Little is known about the nature of most of the variable sources. Only one
source (\#189) is identified with HD\,93343 (O7\,V(n)). The sources
\#36 (=MJ\,156), \#46 (=Tr\,14\,18), \#104
(=DETWC-16\,10,5), \#145 (=Tr\,14\,Y\,66), and \#195
(=DETWC-16\,345) are identified with the 2MASS and optical catalogues but no
spectral type is known. The sources \#74, 80, 115, 137, 177, 194 are
identified with the 2MASS catalogue, no optical identification is known. The
sources \#61, 73, 88, 95, 193 are not identified with any optical or
IR catalogues. All the sources displayed show variability at different
levels, and 3 of them (\#74, 80 and 104) are clearly flaring.

\begin{figure}
\centering
\includegraphics[width=8.5truecm]{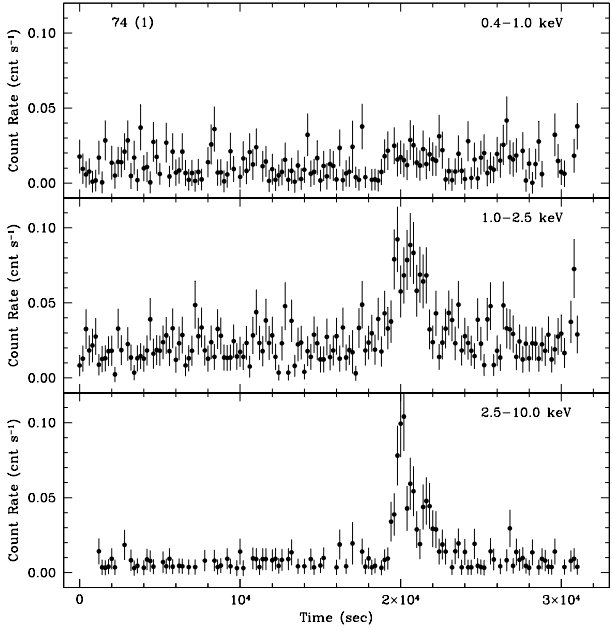}
\caption{{\sc pn} light curves of source \#74 in the soft, medium, and
hard energy bands.}
\label{fig-lc_074}
\end{figure}

\begin{figure}
\centering
\includegraphics[width=8.5truecm]{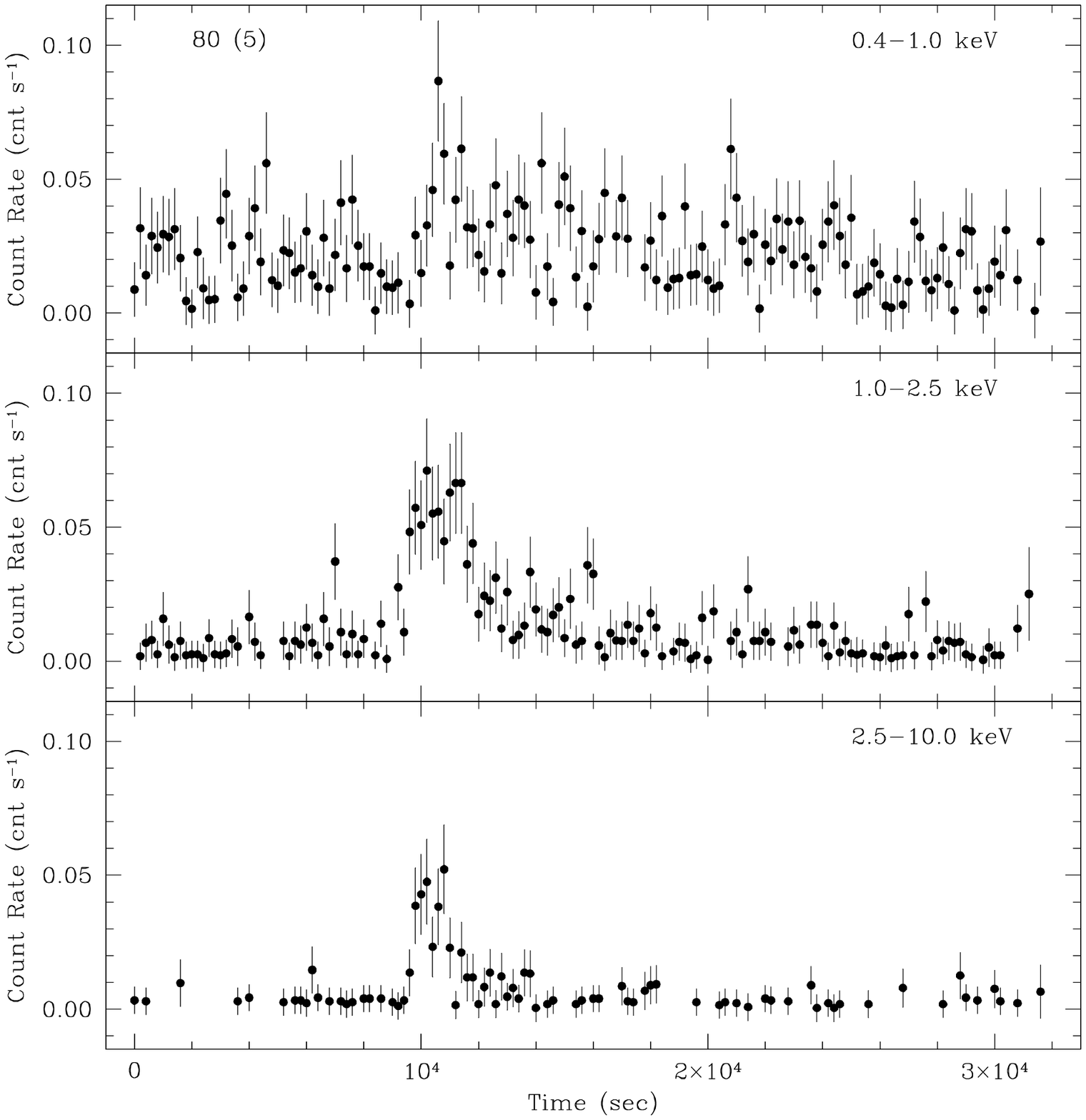}
\caption{{\sc pn} light curves of source \#80 in the soft, medium, and
hard energy bands.}
\label{fig-lc_080}
\end{figure}

\begin{figure}
\centering
\includegraphics[width=8.5truecm]{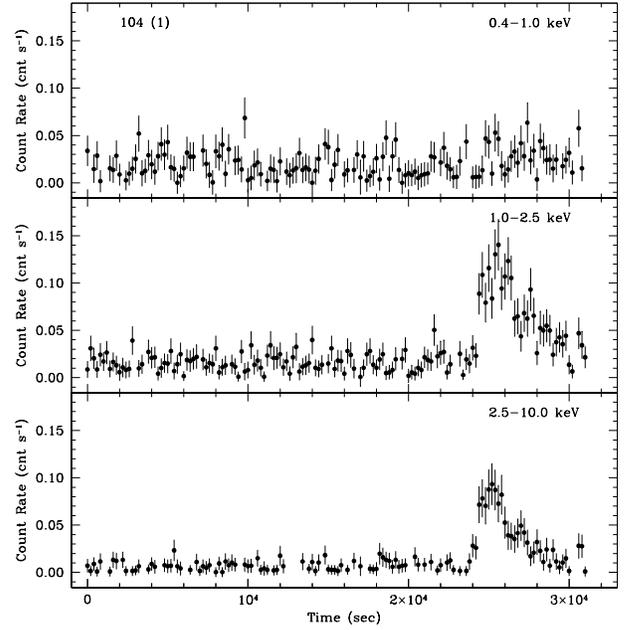}
\caption{{\sc pn} light curves of source \#104 \,=\,
DETWC-16\,10,5 (data set 1) in the soft, medium, and hard energy bands.}
\label{fig-lc_104}
\end{figure}

\subsubsection{The probable pre-main sequence stars \#74, \#80, and
\#104.\label{sect_pms}}

Further insight into the nature of the flare-type sources may be obtained
from the analysis of their light curves and spectra during the flare events.
But first, let us consider the IR properties of the flaring X-ray sources.

\begin{figure}
\begin{center}
\includegraphics[width=8.5truecm]{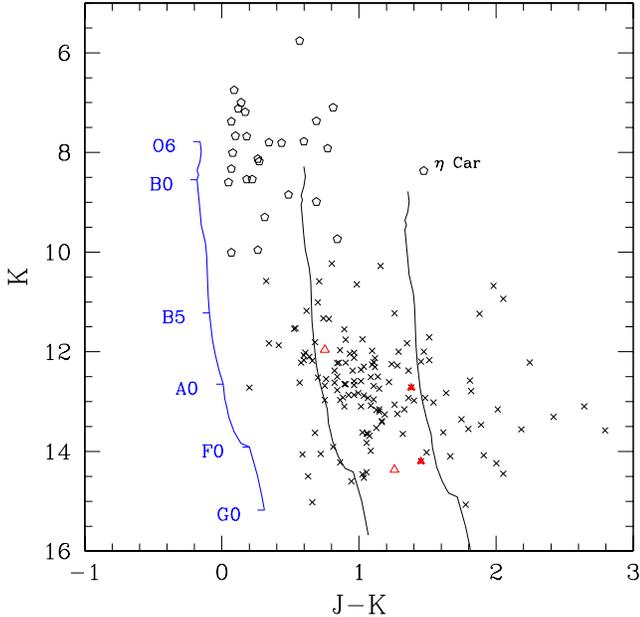}
\end{center}

\caption{$K$ vs.\ $J - K$ colour-magnitude diagram of the 2MASS counterparts
with good quality near-IR photometry of the X-ray sources in the EPIC field
of view. The open pentagons indicate early-type stars, whilst the triangles
stand for flaring X-ray sources. The locus of the un-reddened main sequence
is indicated for a distance modulus of 12.0 by the leftmost solid
line. The reddened main sequence is indicated for two different reddening
values $A_V = 5$ (middle solid line) and $A_V =10$ (rightmost
solid line).}

\label{2masscoulmag}

\end{figure}

Fig.\,\ref{2masscoulmag} illustrates the $K$ vs.\ $J - K$ colour-magnitude
diagram of the 2MASS counterparts in the EPIC field of view. We used the
March 2003 update of the colour transformations, initially derived by
Carpenter (\cite{Carpenter01}) and available on the 2MASS
website\footnote{\tt http://www.ipac.caltech.edu/2mass/index.html}, to
convert the 2MASS colours and magnitudes into the homogenized $JHK$
photometric system introduced by Bessell \& Brett (\cite{Bess88}).

Assuming all the  sources are at the same distance, the sources to the right
of the main sequence could indicate the existence of a population of
pre-main sequence stars. The counterparts of three flaring X-ray sources are
illustrated by the triangles (downward triangles stand for the two
possible counterparts of source \#104, whilst the upward triangles show
sources \#74 and \#80). The flaring X-ray sources with single 2MASS
counterparts are clearly located at positions consistent with these objects
being PMS stars.

The X-ray light curves of the three flaring sources are shown in
Fig.\ref{fig-lc_074}--\ref{fig-lc_104}. It is immediately evident that the
flares occur mostly in the hard energy range which is consistent with
these observed flares being produced by the coronal activity in PMS stars.

Assuming that the plasma producing the observed variabiltiy is confined in a
closed coronal loop as is the case for solar flares, we can use the method
proposed by Serio et al.\ (\cite{Serio91}). These authors established an
analytical relation between the loop half-length $l$, the maximum
temperature at the top of the loop $T_{\rm max}$ and the thermodynamic decay
time $\tau$ (see also Briggs \& Pye \cite{Briggs03}, Giardino et al.\
\cite{Giardino04} and Favata \cite{Favata05}).

\begin{table*}[htb]

\caption{Properties of the flaring sources.}

\label{tbl-flare}

\begin{center}
\begin{tabular}{r c r c c c c c c}
\hline
Source & Rev.\ & \multicolumn{1}{c}{$\tau$} & $kT_{\rm obs}$ & $l \times F(\zeta)$ & $f_X$ (0.4 -- 10\,keV) & $J - K$ & $K$ \\
       &     & \multicolumn{1}{c}{(ksec)} &(keV)   & (R$_{\odot}$) & (erg\,cm$^{-2}$\,s$^{-1}$) &        &     \\
\hline
   74 & 115 & $1.5 \pm 0.3$ & $13.3 \pm 7.5$ & $1.1 \pm 0.3$ & $9.3 \times 10^{-13}$ & 1.452  & 14.198\\
   80 & 285 & $2.5 \pm 0.6$ &  $4.2 \pm 1.1$ & $0.9 \pm 0.2$ & $2.9 \times 10^{-13}$ & 1.381  & 12.718\\
  104 & 115 & $2.1 \pm 0.4$ &  $4.6 \pm 0.7$ & $0.8 \pm 0.1$ & $9.1 \times 10^{-13}$ &        &       \\
\hline
\end{tabular}

\end{center}

\footnotesize
$\tau$ yields the $1/e$ decay time, $kT_{\rm obs}$ is the best fit
plasma temperature during the flare and $l$ is the loop half-length.
The X-ray fluxes correspond to the spectrum during the flare.

\end{table*}

We have determined the $1/e$ decay times of the flares of sources \#74, 80
and 104 by a least square fit of the exponential decay phase of the observed
light curves. Considering the typical count rates of the sources studied
here, the most useful time bins to study the light curves are 500 and 200\,s
for the {\sc mos} and {\sc pn} data respectively. The decay times derived from the fits
of the different EPIC light curves agree within 20\%. The results are given
in Table\,\ref{tbl-flare}. All three sources have rather short decay times
indicating that the X-ray emitting loops must be relatively compact. We
fitted the spectra of the sources during the flare with an absorbed single
temperature {\tt apec} model. The best-fit column densities of the three
sources are found to be $0.26$ -- $0.39 \times 10^{22}$\,cm$^{-2}$. The
best-fit temperatures are quite high ($kT \geq 4.2$\,keV) as expected for
pre-main sequence stars. The measured decay time of the flare actually
exceeds the intrinsic thermodynamical decay time due to the effect of
heating that can extend into the decay phase of the flare. The relation between the two
$F(\zeta) = \tau_{\rm obs}/\tau_{\rm th}$ can in principle be
determined from the slope $\zeta$ of the flare decay in a $\log{T}$ --
$\log{\sqrt{EM}}$ diagram (Favata et al.\ \cite{Favataetal05}).
Unfortunately, the flaring sources discussed here were too faint to allow a
time-resolved spectral analysis. Therefore, we can only
determine $l \times F(\zeta)$. For flaring loops that are freely decaying
with no heating, $F(\zeta) \simeq 2$ (Favata et al.\
\cite{Favata05}) which actually represents the lower limit on $F(\zeta)$.
Therefore, the typical loop sizes of the three sources are
$\leq 0.55$\,R$_{\odot}$ (see Table\,\ref{tbl-flare}).

\subsection{Set-to-set variability}

To detect possible variability of the X-ray sources between the data
sets we performed a $\chi^2$ test at three different confidence
levels (90, 95 and 99\%) to test the null hypothesis of a constant
count rate level in the data. The tests were performed for the count
rates in the soft $S$, medium $M$, hard $H$, and total
(0.4--10.0\,keV) energy bands as well as for the hardness ratios
$HR_1$ and $HR_2$, for all EPIC instruments. The results of the
test are included in Table\,\ref{table_cts} (second column). The
variability status is assigned as follows. If the null hypothesis is
rejected at a confidence level of 95\% in the total energy range in
all instruments simultaneously, the source is labeled ``Var''. If
the null hypothesis cannot be rejected in all instruments, the
source is labeled ``Const''. If the null hypothesis is rejected in
some instruments only, the variability status is set to ``Uncert''.
Finally, if there is not sufficient data to perform the test, the
variability status is set to ``NoInfo''.

Of course, given the limited number of data sets, the detected
variabiltiy of a source tells us hardly more than the mere fact that
the source is variable on a time scale of one day or one year (the
time intervals between our data sets). The nature of this
variability for the majority of the sources remains uncertain. The
purpose of this test was to set a preliminary base for future
studies. Some sources like HD\,93205 (\#117) are early-type
binaries, in which case the X-ray flux varies with the orbital
motion. Spectral analysis may help in disentangling the nature of
the detected sources and their variability. Such an analysis will be
the subject of forthcoming papers.

\section{X-ray properties of early-type stars}

We extracted the spectra of all early-type stars detected in the images,
using the same apertures as for their light curves. The background
spectra were extracted from circular apertures chosen in source-free areas
as close as possible to the sources under investigation, so that the diffuse
background emission is similar to that at a source position. The
corresponding redistribution matrices and ancillary response files were
generated for every source, the spectra were rebinned so that they contain a
minimum of 10 counts per channel.

\begin{table*}

\caption{Parameters of spectral fitting for early type stars in the Carina
Nebula.}

\label{tbl-OB}
\tiny
\begin{tabular}{rcccccccccc}
\hline
      &                            &                         &         &         &                         &                             &                        &               &       \\
  X\# &   Source Name              &    Spectral Type        &$E(B-V)$ &$N^{\rm H}_{\rm ISM}$& $N_{\rm H}$ &          kT                 &        Norm            & $\chi^2/dof$  & dof   \\[+1mm]
      &                            &                         &         &($10^{22}$\,cm$^{-2}$)&($10^{22}$\,cm$^{-2}$)& (keV)             &     ($10^{-4}$)        &               &       \\[+1mm]
\hline\\
   26 & MJ 126                     & B2\,Vb                  & $0.42$  &  $0.24$ &  $0.24_{-0.00}^{+0.06}$ & $0.67_{-0.06}^{+0.06}$      & $0.49_{-0.07}^{+0.12}$ &    1.14       &  135  \\[+1mm]
   30 & CPD-58$^\circ$2611         & O6\,V((f))              & $0.45$  &  $0.26$ &  $0.76_{-0.12}^{+0.11}$ & $0.54_{-0.07}^{+0.07}$      & $0.89_{-0.25}^{+0.29}$ &    1.69       &   51  \\[+1mm]
   33 & Tr 14 21                   & O9\,V                   & $0.61$  &  $0.35$ &  $1.21_{-0.14}^{+0.14}$ & $0.19_{-0.06}^{+0.06}$      & $44.1_{-41.0}^{+41.0}$ &    1.45       &  225  \\[+1mm]
      &                            &                         &         &         &  $0.00_{-0.00}^{+0.38}$ & $2.19_{-0.21}^{+0.22}$      & $2.30_{-0.47}^{+0.47}$ &               &       \\[+1mm]
   45 & HD\,93129\,A+B             & O3\,Iab...+O3.5\,V      & $0.56$  &  $0.32$ &  $0.77_{-0.06}^{+0.07}$ & $0.14_{-0.02}^{+0.01}$      & $392_{-168}^{+870}$    &    1.11       & 1191  \\[+1mm]
      &                            &                         &         &         &  $0.16_{-0.03}^{+0.06}$ & $0.50_{-0.02}^{+0.03}$      & $33.2_{-3.9}^{+3.6}$   &               &       \\[+1mm]
      &                            &                         &         &         &  $0.68_{-0.17}^{+0.22}$ & $1.78_{-0.13}^{+0.13}$      & $11.6_{-1.3}^{+1.4}$   &               &       \\[+1mm]
   46 & Tr 14 18$^\dag$            & B0\,V                   & $0.52$  &  $0.30$ &  $0.30_{-0.00}^{+0.05}$ & $2.43_{-0.20}^{+0.18}$      & $2.36_{-0.12}^{+0.15}$ &    1.22       &  185  \\[+1mm]
   51 & HD\,93130                  & O6\,III(f)              & $0.48$  &  $0.28$ &  $0.28_{-0.00}^{+0.02}$ & $0.58_{-0.02}^{+0.02}$      & $0.68_{-0.04}^{+0.05}$ &    1.13       &  263  \\[+1mm]
   59 & Tr 16 124                  & B1\,V                   & $0.50$  &  $0.29$ &  $0.29_{-0.00}^{+0.06}$ & $2.30_{-0.31}^{+0.42}$      & $0.46_{-0.05}^{+0.05}$ &    1.16       &  101  \\[+1mm]
   63 & HD\,93160                  & O6\,III(f)              & $0.31$  &  $0.18$ &  $0.58_{-0.07}^{+0.09}$ & $0.25_{-0.02}^{+0.03}$      & $6.68_{-1.67}^{+3.38}$ &    0.99       &  258  \\[+1mm]
      &                            &                         &         &         &  $0.35_{-0.35}^{+0.39}$ & $1.25_{-0.24}^{+0.69}$      & $0.66_{-0.21}^{+0.36}$ &               &       \\[+1mm]
   67 & HD\,93161A+B               & (O8\,V+O9\,V)+O6.5\,V(f)& $0.29$  &  $0.17$ &  $0.57_{-0.06}^{+0.08}$ & $0.27_{-0.03}^{+0.02}$      & $7.36_{-1.20}^{+5.63}$ &    1.05       &  342  \\[+1mm]
      &                            &                         &         &         &  $0.61_{-0.32}^{+0.57}$ & $1.16_{-0.24}^{+0.39}$      & $0.81_{-0.22}^{+0.31}$ &               &       \\[+1mm]
   71 & HD\,93162 (WR\,25)$^\dag\dag$ & WN6h                    & $0.69$  &  $0.40$ &  $0.59_{-0.09}^{+0.09}$ & $0.46_{-0.08}^{+0.12}$      & $16.0_{-6.4}^{+4.8}$   &    1.05       & 1017  \\[+1mm]
      &                            &                         &         &         &  $0.29_{-0.10}^{+0.13}$ & $0.75_{-0.03}^{+0.06}$      & $51.6_{-9.6}^{+9.0}$   &               &       \\[+1mm]
      &                            &                         &         &         &  $0.00_{-0.00}^{+0.06}$ & $2.46_{-0.06}^{+0.15}$      & $35.0_{-3.6}^{+2.7}$   &               &       \\[+1mm]
  108 & CPD-58$^\circ$2649         & B0\,V                   & $0.54$  &  $0.31$ &  $0.78_{-0.14}^{+0.11}$ & $0.15_{-0.03}^{+0.06}$      & $21.5_{-19.3}^{+96.6}$ &    1.02       &   92  \\[+1mm]
      &                            &                         &         &         &  $0.00_{-0.00}^{+0.44}$ & $2.39_{-0.65}^{+1.70}$      & $0.24_{-0.05}^{+0.07}$ &               &       \\[+1mm]
  110 & HD\,93204                  & O5\,V((f))              & $0.39$  &  $0.23$ &  $0.34_{-0.11}^{+0.18}$ & $0.18_{-0.03}^{+0.06}$      & $2.44_{-1.70}^{+11.1}$ &    1.18       &  239  \\[+1mm]
      &                            &                         &         &         &  $0.09_{-0.09}^{+0.16}$ & $0.59_{-0.04}^{+0.04}$      & $0.91_{-0.14}^{+0.12}$ &               &       \\[+1mm]
  117 & HD\,93205                  & O3.5\,V+O8\,V           & $0.40$  &  $0.23$ &  $0.42_{-0.15}^{+0.04}$ & $0.18_{-0.01}^{+0.02}$      & $10.8_{-10.4}^{+4.5}$  &    1.13       &  524  \\[+1mm]
      &                            &                         &         &         &  $0.03_{-0.01}^{+0.13}$ & $0.58_{-0.02}^{+0.02}$      & $3.20_{-0.35}^{+0.28}$ &               &       \\[+1mm]
  126 & HDE\,303311                & O5\,V                   & $0.44$  &  $0.26$ &  $0.26_{-0.00}^{+0.09}$ & $0.29_{-0.04}^{+0.02}$      & $0.81_{-0.08}^{+0.69}$ &    1.05       &  238  \\[+1mm]
      &                            &                         &         &         &  $0.00_{-0.00}^{+0.28}$ & $1.82_{-0.40}^{+0.43}$      & $0.27_{-0.04}^{+0.07}$ &               &       \\[+1mm]
  135 & CPD-59$^\circ$2600         & O6\,V((f))              & $0.48$  &  $0.29$ &  $0.29_{-0.00}^{+0.02}$ & $0.60_{-0.03}^{+0.02}$      & $1.01_{-0.26}^{+0.08}$ &    1.11       &  275  \\[+1mm]
      &                            &                         &         &         &  $0.00_{-0.00}^{+2.51}$ & $4.84_{-2.35}^{+23.4}$      & $0.16_{-0.06}^{+0.22}$ &               &       \\[+1mm]
  142 & HD\,93250                  & O3\,V                   & $0.48$  &  $0.29$ &  $0.29_{-0.00}^{+0.02}$ & $0.31_{-0.02}^{+0.02}$      & $4.30_{-0.32}^{+0.73}$ &    1.01       & 1408  \\[+1mm]
      &                            &                         &         &         &  $0.31_{-0.05}^{+0.06}$ & $0.74_{-0.02}^{+0.03}$      & $6.66_{-0.82}^{+1.17}$ &               &       \\[+1mm]
      &                            &                         &         &         &  $0.00_{-0.00}^{+0.03}$ & $3.05_{-0.15}^{+0.21}$      & $6.82_{-0.46}^{+0.36}$ &               &       \\[+1mm]
  147 & CPD-59$^\circ$2603         & O7\,V+O9.5\,V+B0.2\,IV  & $0.43$  &  $0.25$ &  $0.26_{-0.01}^{+0.21}$ & $0.32_{-0.03}^{+0.03}$      & $0.37_{-0.32}^{+0.82}$ &    1.20       &  165  \\[+1mm]
      &                            &                         & $0.43$  &  $0.25$ &  $0.42_{-0.14}^{+0.17}$ & $0.69_{-0.09}^{+0.09}$      & $0.32_{-0.20}^{+0.20}$ &               &       \\[+1mm]
  178 & CPD-59$^\circ$2626         & O7\,V                   & $0.65$  &  $0.38$ &  $0.46_{-0.08}^{+0.50}$ & $0.23_{-0.06}^{+0.06}$      & $1.19_{-1.19}^{+6.60}$ &    1.11       &   55  \\[+1mm]
      &                            &                         &         &         &  $0.63_{-0.34}^{+0.35}$ & $0.84_{-0.20}^{+0.21}$      & $0.79_{-0.59}^{+0.59}$ &               &       \\[+1mm]
  179 & HDE\,303308$^\dag\dag\dag$ & O3\,V((f))              & $0.44$  &  $0.26$ &  $0.60_{-0.09}^{+0.05}$ & $0.12_{-0.01}^{+0.03}$      & $171_{-123}^{+116}$    &    1.13       &  632  \\[+1mm]
      &                            &                         &         &         &  $0.00_{-0.00}^{+0.06}$ & $0.48_{-0.04}^{+0.05}$      & $4.79_{-1.00}^{+0.95}$ &               &       \\[+1mm]
      &                            &                         &         &         &  $3.88_{-0.44}^{+0.49}$ & $3.80_{-0.43}^{+0.44}$      & $0.15_{-0.02}^{+0.02}$ &               &       \\[+1mm]
  183 & CPD-59$^\circ$2629         & O8.5\,V                 & $0.78$  &  $0.45$ &  $0.86_{-0.07}^{+0.06}$ & $0.29_{-0.02}^{+0.03}$      & $9.18_{-2.82}^{+3.43}$ &    0.96       &  924  \\[+1mm]
      &                            &                         &         &         &  $0.15_{-0.13}^{+0.13}$ & $1.92_{-0.12}^{+0.13}$      & $3.16_{-0.23}^{+0.27}$ &               &       \\[+1mm]
  189 & HD\,93343                  & O7\,V(n)                & $0.54$  &  $0.31$ &  $0.63_{-0.12}^{+0.13}$ & $0.29_{-0.05}^{+0.04}$      & $2.45_{-1.14}^{+1.61}$ &    1.33       &  159  \\[+1mm]
      &                            &                         &         &         &  $0.01_{-0.01}^{+1.14}$ & $2.81_{-1.40}^{+1.49}$      & $0.47_{-0.08}^{+0.36}$ &               &       \\[+1mm]
  191 & CPD-59$^\circ$2636         & O7\,V+O8\,V+O9\,V       & $0.56$  &  $0.32$ &  $0.51_{-0.06}^{+0.03}$ & $0.56_{-0.06}^{+0.04}$      & $1.37_{-0.25}^{+0.39}$ &    1.30       &  120  \\[+1mm]
  197 & CPD-59$^\circ$2641         & O6\,V((f))+?            & $0.58$  &  $0.34$ &  $0.54_{-0.07}^{+0.08}$ & $0.64_{-0.05}^{+0.08}$      & $0.95_{-0.23}^{+0.23}$ &    1.12       &  102  \\[+1mm]
  218 & MJ 596                     & O5\,V((f))+O9.5\,V      & $0.73$  &  $0.42$ &  $0.53_{-0.11}^{+0.24}$ & $0.56_{-0.13}^{+0.13}$      & $0.54_{-0.37}^{+0.69}$ &    1.21       &  160  \\[+1mm]
      &                            &                         &         &         &  $0.95_{-0.79}^{+1.64}$ & $2.65_{-0.98}^{+1.15}$      & $1.13_{-0.40}^{+1.02}$ &               &       \\[+1mm]
  235 & HDE\,303304                & O+...                   & $0.61$  &  $0.35$ &  $0.54_{-0.11}^{+0.12}$ & $0.18_{-0.02}^{+0.02}$      & $11.0_{-10.1}^{+17.4}$ &    0.96       &  107  \\[+1mm]
\hline\hline
\end{tabular}

$N^{\rm H}_{\rm ISM}$ is the ISM column density of neutral hydrogen.
Norm is the normalization factor of a thermal model component given by
${\rm Norm}=\frac{10^{-14}}{4{\pi}d^2}{\int}n_en_HdV$, where $d$ is the distance to
the source (cm), $n_e$ and $n_H$ are the electron and H densities
(cm$^{-3}$).

$^\dag$ -- Parameters may be unreliable as the star is located in an area
of extremely varying background and close to a cluster of bright stars.
\newline
$^\dag\dag$ -- Metal abundance $0.59{\pm}0.07$.
\newline
$^\dag\dag\dag$ -- Parameters may be unreliable due to proximity to $\eta$\,Car and a
strong non-uniform background.

\end{table*}

\subsection{Spectral models\label{sect_spectra}}

Table\,\ref{tbl-OB} lists all detected early-type stars along with
the parameters of the spectral fits. Spectral fitting was generally
performed for all data sets and all instruments {\sc mos1}, {\sc
mos2}, and {\sc pn} simultaneously. In some cases when a source
was variable or the data from some instruments in a particular data
set were affected by the CCD gaps, a subset of the data was used. To
quantify the physical properties of the X-ray emission we adjusted a
series of optically thin thermal plasma {\tt apec} models to the
observed spectra using {\tt xspec}. The general form of a
multi-temperature model was {\tt
wabs$_1$*(apec$_1$+wabs$_2$*apec$_2$+...)}. {\tt wabs$_1$} is an
absorption model using the cross-sections for cosmic abundances. We
required that the column density of neutral hydrogen for this model
component was not smaller than the value given in the 5th column of
Table\,\ref{tbl-OB}. The column density was calculated individually
for every star using the empirical relation from Bohlin et al.
(\cite{Bohlin78}). Additional {\tt wabs} components account for
possible internal absorption of X-ray radiation in the winds of
early-type stars. Metal abundances were fixed to the solar value in
all cases except WR\,25, in which case it was a free parameter. In a
few cases an Fe line complex (or indication of it) was present in
the spectra at 6.4-6.8\,keV. This line complex may include the Fe I
fluorescence line at ${\sim}6.4$\,keV and the lines of Fe XXV and
XXVI centered around $6.7$\,keV. The potential presence of the Fe I
line (indicative of cool material) may skew the high temperature
model component required to fit the Fe XXV and XXVI lines. We did a
check for the presence of the Fe I line as follows. We start fitting
with just one thermal model component and include a gaussian
component at $6.4$\,keV. If necessary, we increase the number
of thermal components until a satisfactory fit is achieved. We then
remove the gaussian component and check if the goodness of fit
decreases significantly. If not, we permanently remove this
component from the model. It turned out that no significant Fe I
emission was found in any sources.

Table\,\ref{tbl-OB} does not include the brightest object in the FOV,
$\eta$\,Car. The nature of this star is so complicated that it makes
little sense to give the results of a necessarily superficial analysis here.
A detailed study of {\sl XMM-Newton} observations of this star was
recently published by Hamaguchi et al. (\cite{Hamaguchi07}). A few weak
OB stars were also excluded from the table as no reliable spectral fits
could be performed.

The number of thermal components for different sources varies from one to
three. Spectra of some stars required rather high temperature model
components. Below we will discuss possible reasons for this. We also tried
to fit the spectra with combinations of thermal and power law models. These
did not result in better fits. Note that for some weak sources, the fitted
parameters are not really well constrained and about the same quality of the
fit can be obtained in a single-temperature model for different combinations
of $N_H$ and $kT$. However, as we are most interested in the source {\em
fluxes} within the {\sl XMM-Newton} energy range, the somewhat formal fits
are acceptable. Our model parameters are roughly in agreement with those
obtained by AC03 and EV04 (the latter are based on a Chandra
observation of the region). The exact comparison is impossible as we used
somewhat different models (e.g. different number of temperatures in thermal
models), owing to different quality of the data.

\begin{figure}
\centering
\includegraphics[angle=270,width=8.5truecm]{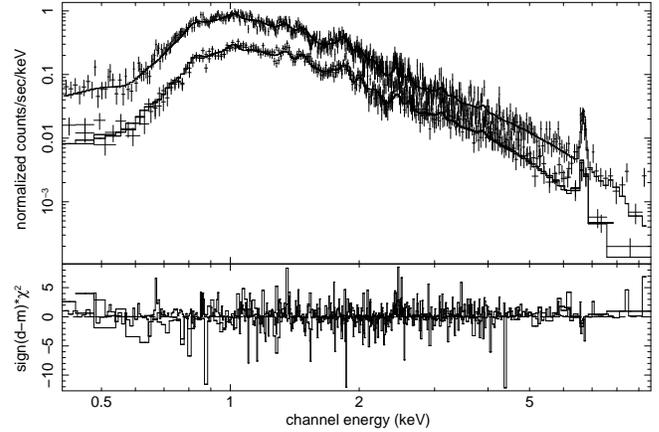}

\caption{EPIC spectra of WR\,25 from the data set 3 along with the best-fit
model. In the top panel, the upper and lower data correspond to the {\sc pn}
and {\sc mos} spectra, respectively, and the solid lines yields the best-fitting
model. The bottom panel shows the contribution of individual energy bins to
the $\chi^2$ of the fit.}

\label{fig-sp_wr25}
\end{figure}

\begin{figure}

\centering
\includegraphics[angle=270,width=8.5truecm]{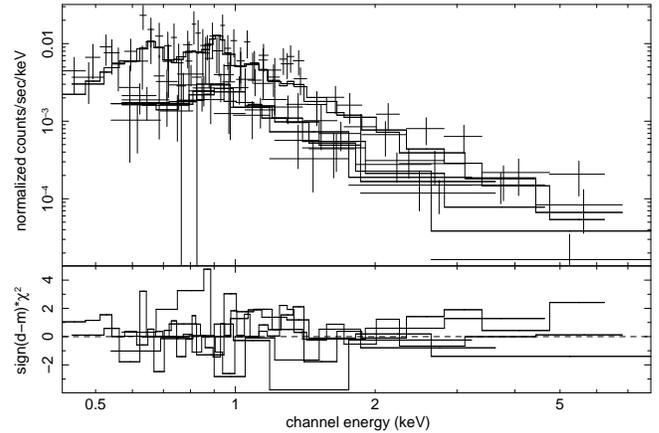}

\caption{Example of a weak source. {\sc pn} spectra of the source \#108
(CPD-58$^\circ$2649) from the data sets 1,2, and 5.}

\label{fig-sp_108}

\end{figure}

\begin{figure}

\centering
\includegraphics[angle=270,width=8.5truecm]{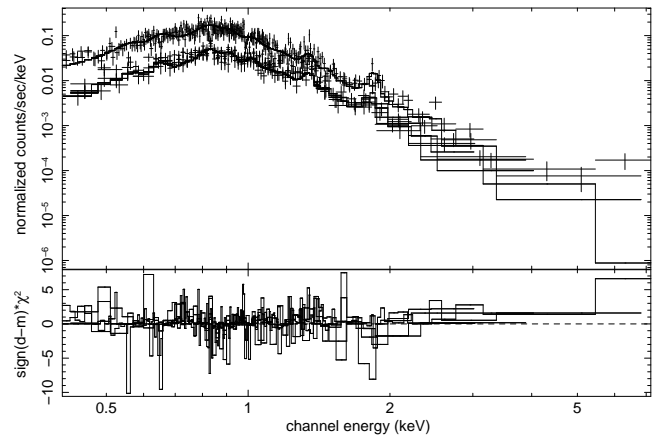}

\caption{{\sc pn} spectrum of the source \#117 (HD\,93205) from the data sets
4 and 5.}

\label{fig-sp_117}

\end{figure}

\begin{figure}

\centering
\includegraphics[angle=270,width=8.5truecm]{5711fig13.eps}

\caption{{\sc pn} spectrum of the source \#142 (HD\,93250) from the data
sets 3--5. Note the presence of the Fe line complex at 6.4--6.7\,keV.}

\label{fig-sp_142}

\end{figure}

A few examples of the observed spectra along with the fitted models are
shown in Fig.\ref{fig-sp_wr25}--\ref{fig-sp_142}. These examples show WR\,25
and some OB stars of different brightness, to illustrate the quality of the
data.

\begin{figure*}
\centering
\includegraphics[width=15.0truecm]{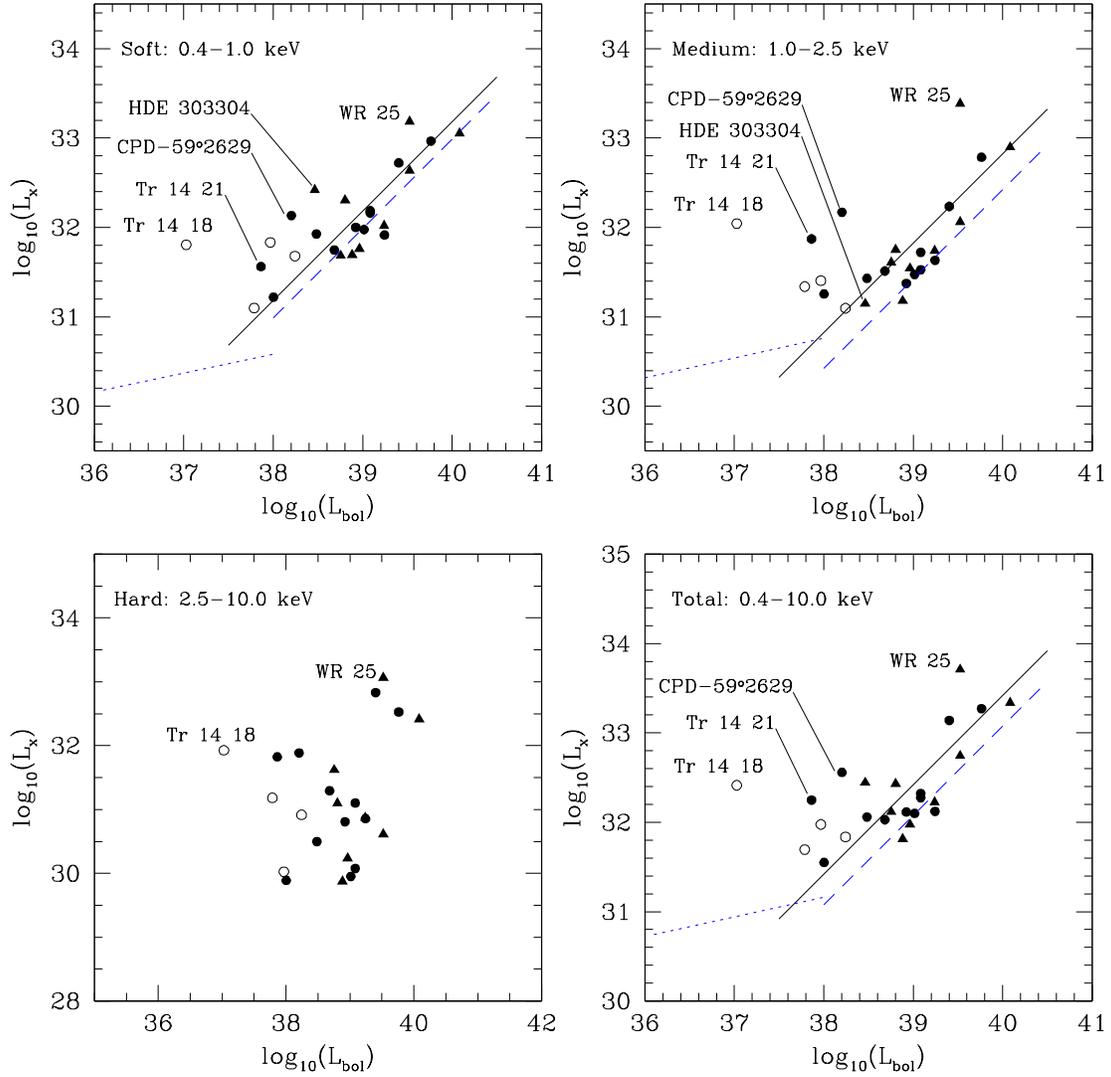}

\caption{$L_{\rm X}$ versus $L_{\rm bol}$ for early-type stars in the Carina
Nebula in different energy bands. B-type stars and WR\,25 are excluded
from the least-square fits. The dashed lines show the best-fit
approximations of the {\em $L_{\rm X}$ -- $L_{\rm bol}$} relation for O (dashed lines)
and B (dotted lines) stars in NGC\,6231 (Sana et al. \cite{Sana06})
(hereafter S06).}

\label{fig-lxlbol}
\end{figure*}

\subsection{X-ray fluxes and L$_x$/L$_{\rm bol}$ ratio}

The X-ray absorbed and unabsorbed (i.e. corrected for the interstellar
absorption) fluxes and bolometric luminosities of OB
stars detected in the {\sl XMM-Newton} data are presented in
Table\,\ref{tbl-LX}. The first column is the source number, the second to
fifth columns are the absorbed X-ray fluxes in the soft, medium, hard, and
total energy bands, while the following four columns are the unabsorbed
fluxes in the same bands. The tenth column is the bolometric magnitude.

EV03 thoroughly discuss bolometric magnitudes for the early type stars they
detected in their Chandra data. We adopted their bolometric magnitudes for
the sources in common. For the stars not detected in the Chandra data, we, like
EV03, compute the bolometric magnitudes using the approach and formulae from
Massey et al. (\cite{MWG00}). Briefly, the reddening $A_{\rm
V}\,=\,R\,\times\,E(B-V)$ with $R\,=\,3.2$ was computed for each star to get
its de-reddened visual magnitude. The bolometric correction of a star was
computed using the relationship $BC\,=\,27.66-6.84\,\times\,\log\,T_{\rm
eff}$. In the complex case of HD\,93161 consisting of three components Aa,
Ab, and B, we computed the bolometric luminosities of every component
separately using the information about the colors and temperatures from
Naz\'e et al. (\cite{Naze05}) and then summed them up to get the total
bolometric luminosity of the system. For MJ\,596 we used the bolometric
magnitude obtained by Niemela et al. (\cite{Niemela06}) from the detailed
analysis of spectral and photometric data.

\begin{table*}

\caption{X-ray fluxes and absolute magnitudes of early type stars in the
Carina Nebula.}

\label{tbl-LX}
\begin{tabular}[h]{rcccc@{\extracolsep{5mm}}cccccccc}
\hline\\
  X\#   & \multicolumn{4}{c}{Absorbed Flux} & \multicolumn{4}{c}{Unabsorbed Flux}  & M$_{\rm bol}$ \\[+1mm]
\cline{2-5}
\cline {6-9}\\

       &[0.4-1.0] & [1.0-2.5] & [2.5-10.0] & [0.4-10.0] & [0.4-1.0] & [1.0-2.5] & [2.5-10.0] & [0.4-10.0] &               \\[+1mm]
\hline\\
  26   &   3.52   &    2.43   &     1.37   &     6.09   &    9.12   &  3.40     &  0.142     &     12.66  &   $-6.21$       \\[+1mm]
  30   &   0.86   &    1.74   &     0.10   &     2.71   &    2.22   &  2.41     &  0.104     &      4.74  &   $-6.30$       \\[+1mm]
  33   &   1.28   &    7.22   &     8.65   &     17.1   &    4.88   &  9.94     &   8.92     &     23.72  &   $-5.95$       \\[+1mm]
  45   &   32.4   &    74.5   &     33.5   &    140.4   &   151.5   & 105.8     &   34.6     &     291.8  &   $-11.5$       \\[+1mm]
  46   &   1.80   &    11.0   &     11.0   &     23.8   &    8.57   & 14.81     &  11.30     &      34.7  &   $-3.86$       \\[+1mm]
  51   &   4.00   &    2.66   &     0.11   &     6.77   &    12.7   &  3.99     &   0.12     &      16.8  &   $-8.83$       \\[+1mm]
  59   &  0.369   &    2.18   &     1.98   &     4.53   &    1.68   &  2.92     &   2.03     &      6.63  &   $-5.76$       \\[+1mm]
  63   &   4.91   &    4.58   &     0.94   &     10.4   &    11.0   &  5.75     &   0.96     &      17.7  &    $-9.4$       \\[+1mm]
  67   &   6.71   &    5.91   &     0.97   &     13.6   &    14.0   &  7.36     &   0.99     &      22.4  &   $-9.39$       \\[+1mm]
  71   &   46.9   &   216.1   &    150.0   &    413.0   &   205.0   & 324.9     &  155.5     &     685.4  &   $-10.1$       \\[+1mm]
 108   &   1.23   &    1.18   &     1.07   &     3.48   &    6.40   &  1.67     &   1.10     &      9.17  &    $-6.9$       \\[+1mm]
 110   &   5.99   &    3.29   &     0.16   &     9.43   &    20.6   &  4.51     &   0.16     &      25.2  &    $-9.0$       \\[+1mm]
 117   &   18.2   &    11.4   &     0.52   &     30.0   &    58.1   &  15.5     &   0.55     &      74.2  &   $-10.1$       \\[+1mm]
 126   &   3.34   &    2.26   &     0.84   &     6.44   &    13.4   &  3.17     &   0.86     &      17.4  &    $-8.6$       \\[+1mm]
 135   &   5.90   &    4.76   &     1.66   &     12.3   &    19.4   &  7.04     &   1.69     &      28.1  &    $-9.0$       \\[+1mm]
 142   &   32.8   &    59.0   &     44.0   &    135.8   &   124.0   &  81.5     &   45.0     &     250.5  &   $-10.7$       \\[+1mm]
 147   &   2.02   &    1.45   &     0.10   &     3.55   &    6.60   &  2.02     &   0.10     &      8.72  &    $-8.5$       \\[+1mm]
 178   &   1.62   &    2.35   &     0.40   &     4.36   &    11.3   &  3.60     &   0.42     &      15.3  &    $-7.5$       \\[+1mm]
 179   &   15.9   &    17.7   &     89.6   &    123.2   &    70.6   &  23.0     &   90.9     &     184.5  &    $-9.8$       \\[+1mm]
 183   &   3.00   &    12.4   &     9.88   &     25.3   &    18.2   &  19.8     &   10.3     &      48.2  &    $-6.8$       \\[+1mm]
 189   &   1.97   &    3.03   &     2.57   &     7.54   &    7.48   &  4.36     &   2.63     &      14.3  &    $-8.0$       \\[+1mm]
 191   &   3.39   &    3.87   &     0.20   &     7.46   &    11.6   &  5.99     &   0.21     &      17.8  &    $-8.3$       \\[+1mm]
 197   &   2.23   &    3.02   &     0.22   &     5.47   &    7.71   &  4.68     &   0.23     &      12.6  &    $-8.7$       \\[+1mm]
 218   &   1.27   &    3.63   &     5.41   &     10.3   &    6.47   &  5.45     &   5.58     &      17.5  &   $-8.18$       \\[+1mm]
 235   &   4.53   &    1.08   &    0.001   &     5.61   &    35.2   &  1.88     & 0.0012     &      37.1  &   $-7.45$       \\[+1mm]
\hline\hline
\end{tabular}

\footnotesize
Fluxes are in units $10^{-14}$\,erg\,cm$^{-2}$\,s$^{-1}$.

\end{table*}

Fig.\,\ref{fig-lxlbol} shows the intrinsic $L_{\rm X}$ versus
$L_{\rm bol}$ for the stars in Table\,\ref{tbl-LX}, in every energy
band. It is known from the literature (e.g. S06) that O- and
B-type stars show a different behaviour. For this reason we marked
them with different symbols. Presumed single O-type stars are marked
with solid circles, binary stars are plotted as solid triangles, and
B-type stars are marked with open circles. The demarcation line
between two types of stars is located at about $L_{\rm bol} =$
$10^{38}$\,erg\,s$^{-1}$. We shall not discuss B-type stars
here as their number in our sample is small.

The X-ray and bolometric luminosities of O-type stars are clearly correlated
in the soft and medium bands. In the hard band the errors are large
so no definite conclusion can be drawn.

The solid lines in Fig.\,\ref{fig-lxlbol} represent the least-square fits of
the observed distributions of O-type stars (both binary and single) in the
form

$$
\log(L_{\rm X}) = a + \log(L_{\rm bol})
$$

\noindent which represents a simple scaling relation $L_{\rm X} =
10^a\,\times\,L_{\rm bol}$. Our data do not allow us to search for a
more sophisticated relation e.g. in the form of a power law.

As seen from Fig.\,\ref{fig-lxlbol}, there is no clear distinction between
binary and single O-type stars in the $L_{\rm X}$\,--\,$L_{\rm bol}$ plane.
We performed various tests on the two samples taken separately. It turned
out that the correlation coefficients for the two samples, the parameters of
the least square fits, and the fit residuals were all very close. The
numerical values of the least-square unweighted fit parameters for the
whole (single plus binary) sample of O-type stars are:

\begin{eqnarray}
\log(L_{\rm X}^{soft})   & = & (-6.82 \pm 0.56) + \log(L_{\rm bol}) \nonumber \\
\log(L_{\rm X}^{med})    & = & (-7.18 \pm 0.92) + \log(L_{\rm bol}) \\
\log(L_{\rm X}^{total})  & = & (-6.58 \pm 0.79) + \log(L_{\rm bol}) \nonumber
\end{eqnarray}

The errors of the parameter are obtained {\em assuming} a good fit
(Press et al. \cite{NR}) and are rather formal. The dispersion of the
luminosities around Eqs.\,1 are $29$\%, $48$\%, and $41$\% in the three
bands above.

In the soft and medium bands, a few O-type stars deviate from the
least-square scaling relation more than the others. These stars are
labeled with their names in Fig.\,\ref{fig-lxlbol}. There is no
unique reason for this deviation. HDE\,303304 shows increased X-ray
luminosity in the soft band. The star is somewhat unusual; it is
classified as ``OB+(le)'' (that is having at least one emission line
in its spectrum) by Stephenson \& Sanduleak (\cite{SS71}). It is not
clear if this can be related to the increased luminosity in the soft
band. On the other hand, the binary hypothesis is not very likely.
If the increased X-ray flux is due to wind-wind collision, then the
\mbox{X-ray} spectrum should also be {\em harder} than a typical
spectrum of a single star, and thus the \mbox{X-ray} limunosities in
the medium and hard \mbox{X-ray} bands ought to be higher than
average. Evidently this is not the case. More likely is that the
soft X-ray flux of HDE\,303304 is affected by the diffuse background
X-ray emission. The source is weak and the diffuse X-ray emission
has maximal flux in the soft X-ray band. It is also very
inhomogeneous across the field of view.

The O8.5\,V star CPD\,-59$^\circ$2629 (=Tr 16 22) is also discovered
as an unusually bright and hard X-ray source for its spectral type
by Chandra instruments (EV04). The latter authors suggested that the
unusually large $L_{\rm X}/L_{\rm bol}$, as well as a surprisingly
hard spectrum, could indicate the presence of a wind confined by the
magnetic fields of this star. Another possible explanation
could be the presence of a low-mass pre-main sequence companion.
These objects can have $L_{\rm X}/L_{\rm bol}$ up to 0.001 and even
0.01 (during flares). However, in the absolute terms, it is unlikely
that they would be bright enough to be seen against the intrinsic
X-ray emission of an early-type star. E.g., In NGC\,6231, PMS stars
can typically display $L_{\rm X}$ up to $10^{31}$\,erg s$^{-1}$ (Sana
et al. \cite{Sana07}).

The deviations in the soft and medium bands of Tr\,14\,21 is likely
related to the fact that this weak source is located near the edge of the
field of view, in an extremely crowded area with strong and variable
background diffuse emission.

Despite the fact that some deviations from the least-square scaling
relations in Fig.\,\ref{fig-lxlbol} could be understood as
instrumental effects we did not exclude the corresponding stars from
the fits. The reason is that, even for stars which are visually
``well-behaved'' , in the plots the uncertainies in the exact value
of the diffuse background emission, and especially in the inferred
neutral hydrogen column densities, may still be important factors
influencing the exact location of the data points in
Fig.\,\ref{fig-lxlbol}. We believe that excluding the deviating
points for the sole purpose of getting a nicer-looking plot is not a
good approach in this case.

\subsection{Comparison with X-ray fluxes from other reviews}

Before we turn to the discussion of the results let us briefly check
the consistency of the fluxes obtained in the current study,
with the previous values presented in EV03 and AC03.

First, note that five OB stars from the list of EV03 (\#8, 9, 39, 44,
141) are not detected in our {\sc XMM-Newton} data, even despite the
fact that the effective telescope area and exposure times are larger in the
{\sl XMM} case. This is explained by the lower spatial resolution of {\sl
XMM}. All these stars are weak X-ray sources. Stars \#8 and \#9 are located near
the brightest object in the region -- $\eta$\,Car (about 39\arcsec and
57\arcsec respectively), and are lost in the bright background created by
the Homunculus. The EV03 source \#44 is between the {\sl XMM} close sources 
\#189 and \#191, the EV03 source \#141 is very weak and is
located near another XMM source \#90. As a consequence, these sources were
not detected by the SAS detection routine.

\begin{figure}
\centering
\includegraphics[width=8.5truecm]{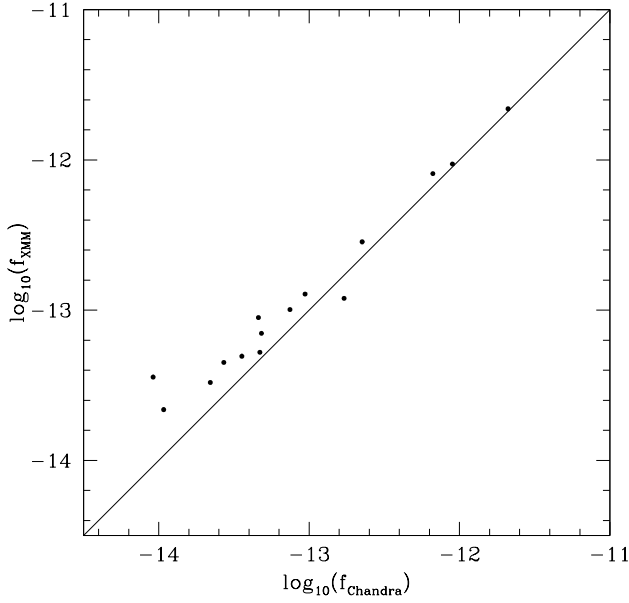}

\caption{Comparison of absorbed {\sl XMM-Newton} and Chandra fluxes in the
energy band 0.5--2.04\,keV.}

\label{fig-XMM_Chandra}
\end{figure}

A comparison of absorbed {\sl XMM} and Chandra fluxes for common
early-type sources is shown in Fig.\,\ref{fig-XMM_Chandra}. EV03
computed their unabsorbed Chandra fluxes in the energy band
0.5--2.04\,keV from the observed count rates assuming a
single-temperature plasma model with the temperature kT=0.385\,keV
and the hydrogen column density
$N_H\,=\,0.3\,\times\,10^{22}$\,cm$^{-2}$. We converted these fluxes
(taken from the Table\,3 of EV03) to the absorbed ones using the
above model parameters. {\sl XMM-Newton} absorbed fluxes were
computed in the same energy band 0.5--2.04\,keV using the best-fit
models from Table\,\ref{tbl-OB}. The results of the two
telescopes are in quite good agreement, especially accounting for
the calibration issues at the time of Chandra observation.

\begin{figure}
\centering
\includegraphics[width=8.5truecm]{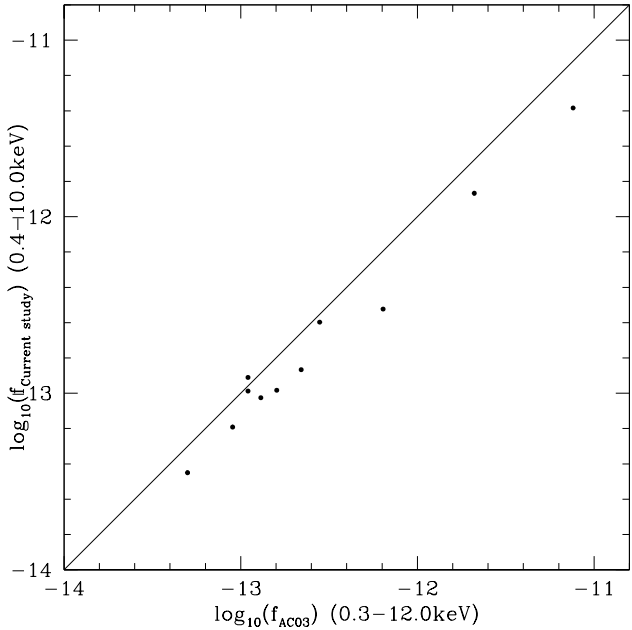}

\caption{Comparison of absorbed fluxes from the current study (energy
band 0.4 -- 10.0\,keV) and the study of AC03 (energy band 0.3 --
12.0\,keV).}

\label{fig-XMM_AC03}
\end{figure}

A comparison of absorbed fluxes of early-type stars from the current
study with the fluxes from AC03 (their Table 3) is shown in
Fig.\,\ref{fig-XMM_AC03}. Two weak stars with problematic fluxes (our
\#179 and \#197) are excluded from the plot. The AC03 fluxes are computed in
the energy range 0.3--12\,keV which explains the systematic shift between
our and their fluxes. Otherwise, the agreement is quite good.

\section{Discussion}

\subsection{General properties of the field}

The X-ray images of the Carina region are dominated by diffuse emission and
strong discrete sources. The brightest discrete sources are
associated with the hot massive stars in the various clusters that populate
this region (see Sect.\,\ref{sect_car_review}). Among the fainter sources,
there are likely some extragalactic background objects, but the vast
majority are likely to be low-mass stars, either foreground coronal sources
or pre-main sequence stars in the Carina complex (see Sect.\
\ref{sect_pms}). Compared to some other open clusters such as NGC\,6231,
there are less PMS candidates detected. This could be due to a lower
detection efficiency (due to the strong diffuse emission) and the larger
distance of the Carina complex compared to NGC\,6231.

\subsection{Early-type stars}

In Section \ref{sect_spectra} we derived the spectral
characteristics and the observed $L_{\rm X}$\,--\,$L_{\rm bol}$
relations for O-type stars in our sample. Recently, a similar
investigation was pulished regarding the early-type population in
NGC\,6231 by S06. The authors also provide a review of the previous
works on this subject. The S06 linear fits of the $L_{\rm
X}$\,--\,$L_{\rm bol}$ relations for O and B stars in NGC\,6231 are
shown in Fig.\,\ref{fig-lxlbol} by dashed and dotted lines
respectively.

When comparing our results with those of S06 one should keep in
mind that the Carina Nebula is known as a region with very patchy
interstellar absorption, thanks to numerous arms of dust and gas
penetrating through the volume of the stellar associations. For this
reason it is extremely difficult to account accurately for the ISM
absorption. The situation with NGC\,6231 is less complex and this is
a probable explanation for the scatter of data points in our
Fig.\,\ref{fig-lxlbol} which is somewhat larger than in the case of
S06\footnote{The X-ray bands of S06 are almost identical to ours.}.
In addition, the difficulty with the ISM absorption may also
influence the accuracy of the bolometric luminosities of the stars
under investigation.

The numerical parameters of the S06 scaling relations are generally similar
to ours: e.g. for their full sample of O-type stars and in the total energy
band 0.4\,--\,10.0\,keV their $\rm{log}(L_{\rm X}/L_{\rm
bol})\,=\,-6.865\,\pm\,0.186$ (cf. our value for the 0.4\,--\,10.0\,keV
band: $-6.58\,\pm\,0.79$).

The biggest difference between Sana et al. and our values of the scaling
parameter for O-type stars is in the medium band. It is explained by the
fact that in our fits we included all O-type stars while S06 provide
the best fit value for a sub-sample of O-type stars excluding two deviating
binary systems. If we exclude two anomalous stars Tr\,14\,21 and
CPD$-59^\circ$\,2629 from the medium band fit, the scaling parameter changes
from $-7.18$ to $-7.32$, a value closer to the one of Sana et al.
($-7.58\,\pm\,0.2$). Our results on the $L_{\rm X}$\,--\,$L_{\rm bol}$
relation can be summarised as follows:

\begin{enumerate}

\item The canonical relation $L_{\rm X}\,\sim\,10^{-7}\,\times\,L_{\rm bol}$
is mainly defined by relatively soft X-ray fluxes below 2.5\,keV. In
the hard band (E $> 2.5$\,keV), the measured X-ray flux is
probably poorly constrained, except for the brightest sources. As a
consequence, the scatter of the hard X-ray luminosities is large and
no definitive fit possible.

\item S06, based on uniform X-ray data and homogeneous observations, state
that the observed dependence between intrinsic $L_{\rm X}$ and $L_{\rm bol}$
is much tighter than was thought before, with the residual dispersion of the
$L_{\rm X}$\,--\,$L_{\rm bol}$ relation around the least-square scaling law
being less or equal to some 20\%, if one excludes binary stars and stars later than
O9. Our current results are unable to confirm this statement due to a
larger scatter in the data owing to the patchy ISM absorption in the Carina
region.

\item {\em On average}, the {\em observed} X-ray properties of O
binary stars are not very different from those of single stars (cf. our
Fig.\,\ref{fig-lxlbol} and Fig.\,16 from S06). We shall discuss
possible reasons for this behaviour below.

\end{enumerate}

\subsection{On the origin of the $L_{\rm X}$\,--\,$L_{\rm bol}$ scaling for
single stars}

It has been pointed out many times in the literature that the scaling
relation between $L_{\rm X}$ and $L_{\rm bol}$ could be an indirect effect.
Indeed, the common paradigm for the generation of X-rays in early-type stars
is a wind shock model (see e.g. Lucy \& Solomon \cite{Lucy70}, Owocki \&
Cohen \cite{OwockiC99}, Dessart \& Owocki \cite{Dess05}). If X-ray emission
is formed throughout the wind then $L_{\rm X}$ should be proportional to
some combination of wind parameters like the mass loss rate $\dot{M}$ and
the terminal wind velocity $v_\infty$. Indeed, Owocki \& Cohen
(\cite{OwockiC99}) from the exospheric approximation established that the
X-ray luminosity should scale with the mean wind density $\dot{M}/v_\infty$
in the form $L_{\rm X}\,\sim\,(\dot{M}/v_\infty)^2$ for optically thin winds
and $L_{\rm X}\,\sim\,(\dot{M}/v_\infty)^{(1+s)}$ for optically thick winds,
with $s$ being the index of the radial power law dependence of the X-ray
filling factor $f\,\sim\,r^s$. However, Sana et al. (\cite{Sana06}) state
that these considerations contrast with the observed tight dependence
between $L_{\rm X}$ and $L_{\rm bol}$.

A new idea about the origin of X-rays from single O-type stars was
recently brought forward by Pollock (\cite{Pollock07}). It is based
on the fact that, according to recent {\sl XMM-Newton} high
resolution observations of $\zeta$\,Orionis, all emission lines have
the same velocity profile which contrast with the paradigm of wind
shocks, in which line profiles should be dependent on the global
kinematic structure of the wind. The principal difference between
the two models is that in the traditional wind shock model the
energy exchange between post-shock ions and electrons occurs on a
short time scale while Pollock estimates the ion-ion collisional
mean-free path and find that at least in the case of
$\zeta$\,Orionis it can be quite large, which leads to very slow
energy exchange between hot ions and cold electrons. He further
suggest that the shocks are collisionless and that the ionization of
the post-shock gas is not caused by electrons as in the traditional
picture but rather by protons. The line profiles then are defined
not by macroscopic motion of ionized gas but simply by the
line-of-sight component of the thermalized motion of ions in the
immediate post-shock gas. This idea has not yet been developed up to
a stage when numerical modeling could be done and a theoretical
connection between $L_{\rm X}$ and $L_{\rm bol}$ (if any)
established.

From the observational point of view, we agree with S06 that
whatever the theoretical explanation behind the $L_{\rm
X}$\,--\,$L_{\rm bol}$ relation, it is a firmly established
observational fact which should be explained by a future theory. Not
only is the proportionality itself now well established but also
some of its more detailed characteristics (see items 1\,--\,3
above).

\subsection{Binary versus single stars}

Item (3) in the above list requires special consideration. It was
already mentioned in the introduction that, in a binary system
consisting of two early-type stars, the supersonic winds of the
components will inevitably collide. The key feature of the wind-wind
collision, making it different from localized shocks formed by
internal line-driven instability of supersonic winds is that the
relative velocity of the two flows is very high, of the order of
thousands km\,s$^{-1}$, at least along the line connecting the
centers of the components. For comparison, the velocity differences
in localized shocks within a wind are typically of the order of a
few hundred km\,s$^{-1}$. As a result, X-ray emission from binary
early-type stars should be more luminous and {\em harder} than in
single stars. This was understood long ago (Prilutskii \& Usov
\cite{Pril76}, Cherepashchuk \cite{Cher76}). One of the first
theoretical predictions about the hardness and X-ray luminosity of
binary early-type stars was made by Usov (\cite{Usov92}) and later
on was refined and quantified in numerical gas-dynamical simulations
(see, e.g. Stevens \& Pittard \cite{Stev99}).

Over the years, it became clear that the original somewhat {\em
naive} expectations of more luminous and hard X-ray radiation from
colliding wind binaries had to be significantly adjusted. A large
number of factors play a role in the X-ray output of a binary, such
as the orbital period (hence the distance between the components,
the pre-shock velocity and the density in the collision zone), the
wind momentum ratio, the orbital inclination, the eccentricity, etc.
These factors influence both the rate of intrinsic X-ray production
in the collision zone and the absorption of the X-rays within the
winds of the components. For additional discussion, see a
recent paper by Linder et al. (\cite{Linder06}).

It is nevertheless still puzzling that so few early-type binaries are
significantly more luminous and harder than the total sample of
O-type stars (e.g. in our Fig.\,\ref{fig-lxlbol} and in Fig.\,16 of S06,
also see Raassen et al., \cite{Raas03} who present a collection of
X-ray properties of single and binary early-type stars). Two intriguing
questions already mentioned in the introduction are (i)\,why some known
colliding wind binaries have soft X-ray spectra and (ii)\,why some stars
having hard X-ray spectra (as demonstrated e.g. by the presence of the Fe
XXV--XXVI lines at ${\sim}$6.7\,keV) lack any evidence of binarity?

The answer to the first question may be two-fold. First, it may be partly
the problem of the detection of the hard X-ray spectral tail in relatively
faint colliding wind binaries. On the other hand, theoretical studies of
wind-wind collision mechanisms (Stevens \& Pollock \cite{SP94}, Owocki \&
Gayley \cite{Owocki95}, Walder \& Folini \cite{Walder03}, Antokhin et al.
\cite{Ant04}) suggest that either the so-called radiative braking or
radiative inhibition (mutual radiative influence of the binary components on
the wind of their {\em vis-\'{a}-vis}) may significantly slow down the winds
of the binary components just before the collision, thus reducing the X-ray
luminosity of the system and the hardness of its X-ray spectrum (which is
proportional to the square of the wind velocity entering the collision
zone). To verify this possibility, we shall apply the recent models
accounting for this phenomenon, in our forthcoming analysis of the binary
HD\,93205 which has one of the best X-ray coverages of the orbital cycle of
any early-type binary.

There is probably no single answer to the second question, either.
In some cases (most notably WR\,25 and MJ\,596) it was recently
found (see references above) that these hard X-ray sources are
indeed binary systems that escaped detection before. These recent
advances raise interest in a similar case of a bright O3\,V star
HD\,93250 (showing clear indication of the Fe XXV-XXVI lines).
The star does not display any radial velocity or photometric
variability and as such was considered as single. However, it was
reported to display non-thermal radio emission (Leitherer et al.
\cite{Leith95}) and this feature is nowadays often believed to be
evidence for a colliding wind system also in O+O binaries. On the other
hand, the O8.5\,V star CPD\,-59$^\circ$2629 which is also an unusually
bright and hard X-ray source for its spectral type might be an example of a
star with a magnetically confined wind (ud-Doula et al. \cite{Doula06}).
A low-mass companion with high activity could also be considered.

\section{Conclusions}

High quality and large amounts of {\sl XMM-Newton} data obtained for
the Carina Nebula, allowed us to detect 235 discrete sources
in the field of view. Several of these sources are probably pre-main
sequence stars with a characteristic short-term variability,
while seven sources are recognized as possible background AGNs.

We concentrated in our study on the properties of early-type stars in the
region. Spectral analyses of twenty three sources of type OB and WR\,25
were performed. We derived the spectral parameters of the sources and their
fluxes in three energy bands. Estimating a value of interstellar absorption
towards every source and assuming a distance of 2.5\,kpc to the Nebula, we
derived the X-ray luminosities of these stars. Comparison of these luminosities
with bolometric luminosities has confirmed the known scaling relation
$L_{\rm X}\,\sim\,10^{-7}\,\times\,L_{\rm bol}$ for early type stars and
allowed us to make a few more detailed statements about the characteristics
of this relation. We pointed out that, {\em on average}, the observed X-ray
properties of binary and single early type stars are similar, and give
several possible explanations for this fact.

Further work will be aimed at the detailed analysis of the data of
particular sources.

\begin{acknowledgements}

We acknowledge our anonymous referee for numerous comments that
helped to improve this work. We are grateful to the calibration
teams of the instruments on board {\sl XMM-Newton}. The SRON
National Institute for Space Research is supported financially by
NWO. IIA thanks H.Sana for providing very useful SAS scripts for
extracting light curves and spectra of objects. IIA acknowledges
support from the Russian Foundation for Basic Research through the
grant 05-02-17489 and from the Russian LSS project 5218.2006.2. The
Li\`ege team acknowledges support from the Fonds National de la
Recherche Scientifique (Belgium) and from the PRODEX {\sl XMM} and
{\sl Integral} contracts. This research has made use of the
Digitized Sky Survey produced by STScI, and the SIMBAD database,
operated at CDS, Strasbourg, France. This publication makes use of
data products from the Two Micron All Sky Survey, which is a joint
project of the University of Massachusetts and the Infrared
Processing and Analysis Center/California Institute of Technology,
funded by the National Aeronautics and Space Administration and the
National Science Foundation. JCB and IA acknowledge the support of a
PPARC/STFC Rolling Grant

\end{acknowledgements}

{}

\end{document}